# GastroViT: A Vision Transformer Based Ensemble Learning Approach for Gastrointestinal Disease Classification with Grad CAM & SHAP Visualization

Sumaiya Tabassum[1], Md. Faysal Ahamed[1], Hafsa Binte Kibria[1], Md. Nahiduzzaman[1],
Julfikar Haider[2] Muhammad E. H. Chowdhury[3] Muhammad Tariqul Islam[4]

[1]Rajshahi University of Engineering & Technology, Rajshahi-6204, Bangladesh
[2]Department of Engineering, Manchester Metropolitan University, Chester Street, Manchester M1 5GD, UK
[3]Qatar University
[4]Universiti Kebangsaan Malaysia

**ABSTRACT**
The gastrointestinal (GI) tract of humans can have a wide variety of aberrant mucosal abnormality findings, ranging from mild irritations to extremely fatal illnesses. Prompt identification of gastrointestinal disorders greatly contributes to arresting the progression of the illness and improving therapeutic outcomes. This paper presents an ensemble of pre-trained vision transformers (ViTs) for accurately classifying endoscopic images of the GI tract to categorize gastrointestinal problems and illnesses. ViTs, attention-based deep neural networks, have revolutionized image recognition by leveraging the transformative power of the transformer architecture, achieving state-of-the-art (SOTA) performance across various visual tasks. The proposed model was evaluated on the publicly available HyperKvasir dataset with 10,662 images of 23 different GI diseases for the purpose of identifying GI tract diseases. An ensemble method is proposed utilizing the predictions of two pre-trained models, MobileViT_XS and MobileViT_V2_200, which achieved accuracies of 90.57% and 90.48%, respectively. All the individual models are outperformed by the ensemble model, GastroViT, with an average precision, recall, F1 score, and accuracy of 69±0.36%, 63±0.39%, 64±0.38%, and 91.98%, respectively, in the first testing that involves 23 classes. The model comprises only 20 million (M) parameters, even without data augmentation and despite the highly imbalanced dataset. For the second testing with 16 classes, the scores are even higher, with average precision, recall, F1 score, and accuracy of 87±0.13%, 86±0.15%, 87±0.15%, and 92.70%, respectively. This research addresses a method to integrate both local and global features using vision transformers whereas previous works worked mostly with CNN based architectures which rely on capturing local features. The real-time analytical capabilities of the model are validated through comparisons with SOTA models and selected transfer learning (TL) models where by using SOTA models gives accuracy of about 86.60% and the proposed ensemble model gives an accuracy of 91.98%. Additionally, the incorporation of explainable AI (XAI) methods such as Grad-CAM (Gradient Weighted Class Activation Mapping) and SHAP (Shapley Additive Explanations) enhances model interpretability, providing valuable insights for reliable GI diagnosis in real-world settings.

## I. INTRODUCTION

The human gastrointestinal system is made up of a hollow, muscular tube that extends from the mouth cavity to the rectum and anus via the pharynx, esophagus, stomach, and intestines. It usually has two sections, the upper and lower tracts, and is approximately nine meters long[1]. The primary functions of the gastrointestinal tract are food digestion, nutrient absorption, and waste elimination. These activities cannot be effectively performed in the presence of any gastrointestinal diseases. Constipation, irritable bowel syndrome, hemorrhoids, anal fissures, anorectal abscesses, anal fistulas, diverticular diseases, colitis, colon polyps, and colorectal cancer are only a few examples of digestive problems. The World Health Organization (WHO) estimates that 1.8 million individuals worldwide pass away from digestive illnesses each year [2]. The third most common disease in the world and a major cause of death is colorectal cancer (CRC)[3], which usually arises from sporadic colorectal adenomatous polyps. The death rate from this illness can be drastically decreased with early identification. Colonoscopy is a well-recognized procedure for the diagnosis and excision of these lesions, which has been demonstrated to lower the incidence and death from CRC[4]. Identification of the disease is significantly promoted by the endoscopic images[5] that are acquired during the physical examination. Medical professionals like endoscopists, gastroenterologists, and other medical practitioners are in great demand, but they can be costly, labor-intensive, and prone to errors. A technology called computer-aided diagnosis (CAD) makes use of medical image processing and artificial intelligence to assist radiologists in diagnosing diseases by interpreting images.

The likelihood of developing colorectal cancer can increase due to a number of gastrointestinal conditions, including ulcerative colitis, inflammatory bowel disease (IBD), and Crohn's disease. Two distinct approaches have been suggested to minimize the disease burden: early diagnosis and CRC screening in symptomatic patients[6]. Ten percent of all new cancer cases worldwide are CRC, with a higher incidence rate in developed nations [7].A global survey on the illness indicates that 26% of men and 11% of women have been diagnosed with colorectal cancer [8]. According to a global survey conducted in 2017, gastrointestinal disorders claimed 765,000 lives, with colon cancer accounting for 525,000[9]. In 2018, there were over 3.4 million deaths worldwide from GI malignancies, accounting for approximately 4.8 million new cases[10]. In 2021, there were 338,090 new cases of colorectal cancer in the US, and the number of fatal cases increased by 44% [11].

Endoscopy, encompassing both colonoscopy and esophagogastroduodenoscopy (EGD), is the most efficient and effective method for screening the upper and lower intestines for any health issues. Moreover, other critical methods for the comprehensive diagnosis of GI illnesses include capsule endoscopy, endoscopic ultrasonography (EUS), CT scan, magnetic resonance imaging (MRI), and positron emission tomography (PET) scan. The stomach was previously very difficult to examine with a regular endoscopic procedure, but now medical professionals can thank a novel procedure called wireless capsule endoscopy (WCE)[12]. Many optical technologies have been developed in recent years to help endoscopists in diagnosing through the histopathological route. However, these technologies demand the use of highly skilled endoscopists with high-magnification endoscopic equipment. To overcome these restrictions, narrow band imaging (NBI) endoscopes, with or without magnification, can be used in conjunction with the worldwide Narrow Band Imaging Colorectal Endoscopic Classification (NICE) classification. More recently, there has been an increasing interest in the development of computer-aided diagnosis (CAD) systems for automatic polyp detection and histology prediction without polyp excision or biopsy, although these techniques still require highly skilled endoscopists. CAD systems based on machine learning (ML) [13] have been used to successfully classify images of colorectal tumors. These types of ML-based systems require a high level of knowledge and extensive preprocessing of the image datasets. In this case, the superior image classification performance of deep learning (DL[14] over earlier state-of-the-art methods has attracted significant interest in the field of medical image analysis. Transformers have been used for machine translation recently and have demonstrated remarkable promise in a variety of natural language processing applications. The transformer's ability to operate self-attention mechanisms to handle global features motivates it to address DL's shortcomings. ViTs[15] could be an alternative option for GI disease identification and classification due to their potential to improve efficiency and accuracy through sophisticated image processing and pattern recognition.ViTs outperform typical convolutional neural networks (CNNs) in GI image processing due to their ability to perceive global context through self-attention mechanisms. Unlike CNNs, which focus on local features using fixed receptive fields, ViTs process picture patches to learn long-range dependencies over the entire image.This enables scientists to capture diffuse patterns, the entire anatomical context of lesions, and complex interactions between distant parts of the GI tract, resulting in more extensive and robust diagnostic findings. While CNNs excel at local texture analysis, ViTs' capacity to model global relationships makes them especially good for complicated GI image processing where holistic knowledge is critical, often obtaining higher performance with adequate training data.The aim of this work is to effectively classify GI diseases using ViTs. In this work, we address key research gaps in gastrointestinal (GI) disease classification by overcoming the limitations of CNN-based methods, which, despite strong local feature extraction, struggle to capture global contextual information. The generalization and robustness of existing research are limited since they mostly concentrate on local textures and fail to sufficiently address the heterogeneity across various GI organs. To tackle this, we propose GastroViT, an ensemble of two lightweight pretrained vision transformers MobileViT_XS(MViT_XS) and MobileViTv2_200(MViTv2_200), which leverage self-attention to jointly model local and global features, enabling better handling of organ variability in structure, texture, and appearance. Unlike prior works relying solely on CNNs or basic ensembles, our method demonstrates superior performance on a challenging 23-class GI disease dataset, while remaining efficient with only 20M parameters. Moreover, we integrate XAI techniques (Grad-CAM and SHAP) to enhance interpretability and clinical trust by highlighting medically relevant regions. Overall, GastroViT offers a robust, accurate, and interpretable framework for real-time or near real-time GI disease classification, specifically tailored to address the heterogeneity of endoscopic images.

- The study proposes a vision transformer-based ensemble model, GastroViT, which addresses CNN's shortcomings in GI image processing by using its self-attention processes to handle global characteristics.

- The proposed framework is associated with two lightweight pretrained vision transformers where average ensemble of prediction from MViT_XS and MViTv2_200 is employed to classify GI diseases of up to 23 classes with superior performance.
- The class wise performance of the proposed GastroViT model is evaluated through computational resource and time analysis and by comparing with Transfer Learning (TL) models and SOTA models proposed in the literature.
- The interpretability of the proposed framework has been assessed with the integration of explainable XAI such as Grad-CAM and SHAP.

## II. Related Works

There has been a long-cherished desire for the diagnosis of GI tract disorders and polyps to be handled automatically by artificial intelligence techniques. For this reason, and to the extent that advancements in technology permit, numerous studies have been conducted in the past. Furthermore, research on multi-class categorization has been carried out, but in most of cases with a limited number of classes [16]. To identify and examine a range of GI disorders from endoscopic images, researchers have used a number of ML and DL techniques [17]. GI tract disorders were categorized by Yogapriya et al. [18] using pre-trained models, including VGG16, ResNet-18, and GoogleNet. A KVASIR v2 dataset containing 6702 images from eight classes was used by the authors. The VGG16 model achieved the highest accuracy of 96.33%. Several endoscopic images of GI diseases were obtained by Montalbo et al.[19] using a multi-fused residual convolutional neural network (MFuRe-CNN) with auxiliary fusing layers and a fusion residual block, both with alpha dropouts. This MFuRe-CNN model enhanced diagnostic performance and helped to gain robust features by combining three state-of-the-art models into a single feature extraction pipeline with partially frozen and truncated layers. With only 4.8 M parameters, the proposed model achieved an excellent test accuracy of 97.25%. Oztürk et al. [20] presented a long short-term memory (LSTM) layer in the backbone CNN that receives characteristics from each pooling layer to improve the classification's robustness, especially for unbalanced datasets. The final classification was achieved by combining the outputs of all the LSTM layers. Khan et al.[5] suggested wireless capsule endoscopy (WCE) to provide a novel deep learning-based approach for ulcer detection and classification of GI illnesses (ulcer, polyp, hemorrhage). For ulcer segmentation, the method used a modified mask recurrent convolutional neural network (RCNN), which was trained using images annotated as ulcer. The pre-trained CNN model for classification, ResNet101, was refined by transfer learning to extract deep features. Using ResNet50 + FPN as the backbone, the MOC is 0.8807, and the average precision is 1.0. Secondly, the use of the cubic SVM for K=10 achieved the highest classification accuracy of 99.13%. Jain et al.'s [21] sophisticated deep CNN, WCENet, was the best method for identifying and locating anomalies in images that came from WCE. WCENet outperformed other popular models and showed great potential for clinical use with an accuracy of 98% on the KID dataset. Hatami et al. [22] presented a deep learning network for the detection and classification of stomach and precancerous illnesses. This network used excitation and squeeze techniques to improve performance. Lan et al. [23] proposed a hybrid approach that combines a poorly supervised cross-modal embedding framework with unsupervised deep learning approaches. They employed a range of networks, such as long short-term memory (LSTM) and autoencoders, to enable medical practitioners to thoroughly view WCE films. A cutting-edge approach of classification for GI disorders was introduced by Alhajlah et al. [24]. Models such as ResNet-50 and ResNet-152 produced an impressive 96.43% classification performance. Mohapatra et al. [25] proposed CNN and discrete wavelength transformation (WT) methods to classify polyp and esophagitis classes which acquired an accuracy of 96.65%. The latest study by Md. F. Ahamed et al.[26] achieves an accuracy of 87.75% using the GastroVision dataset with 8,000 images of 27 GI diseases, leveraging the proposed Parallel Depthwise Separable Convolutional Neural Network (PD-CNN) as a lightweight feature extractor and Pearson Correlation Coefficient (PCC) for feature selection in a computer-aided diagnosis (CAD) system.

Based on the earlier work, the following shortcomings and challenges are identified. In the earlier work, only a few researchers attained a higher degree of accuracy [27]. Several researchers achieved accuracy rates of 93% to 98% in their research. The researchers used datasets with a limited number of classes. Most authors used a small number of image samples (varying from 4000 to 8000) to evaluate the effectiveness of their suggested models. It is crucial to acknowledge that in order to illustrate the effectiveness of the suggested models, the researchers used datasets with a limited number of classes (between 5 and 8) and a modest number of image samples (between 4000 and 8000). Moreover, the majority of the researchers have used TL- and DL-based models including GIT-NET, MobileNetV2, ResNet-50, ResNet152,EfficientNet [28], mostly CNN-based pre-trained models. These models have shortcomings such as inability to capture long-range dependencies and global context from the beginning. In this context, transformers can scale well with data and computational resources and global features by self-attention mechanism. Furthermore, a significant issue appears with the computing requirements (e.g., size, number of parameters and layers) that result in longer processing times, making it challenging to use the classification model efficiently. Reducing parameter counts, size, and improving other aspects of the existing models is essential to develop new GI disease categorization model. Earlier, a lot of research has concentrated on creating and using machine learning models without including XAI methodologies. XAI improves the model's transparency and interpretability and offers insightful information about the underlying decision-making process. Therefore, for GI disease classification models to be widely used, they must be developed for a large number of multi-class classifications with high accuracy, reduced computing resources, faster processing, and appropriate model interpretation.

## III. Methodology

### A. OVERALL FRAMEWORK

In Figure 1, the proposed methodology is presented. The annotated dataset, which included 23 different GI disease categories. After proceeding through a number of preprocessing stages, the images were chosen for the training and validation sets and resized to 224×224 pixels with three color channels. Advanced transfer learning techniques such as ResNet50, EfficientNetB7, DenseNet201, VGG16, which use standard CNN as their foundation, were used to extract features from the images. Furthermore, advanced models based on Vision Transformers, including CoAtNet0, ResNeXt50, EfficientViT_B2, MViT_XS, and MViTv2_200 were used.

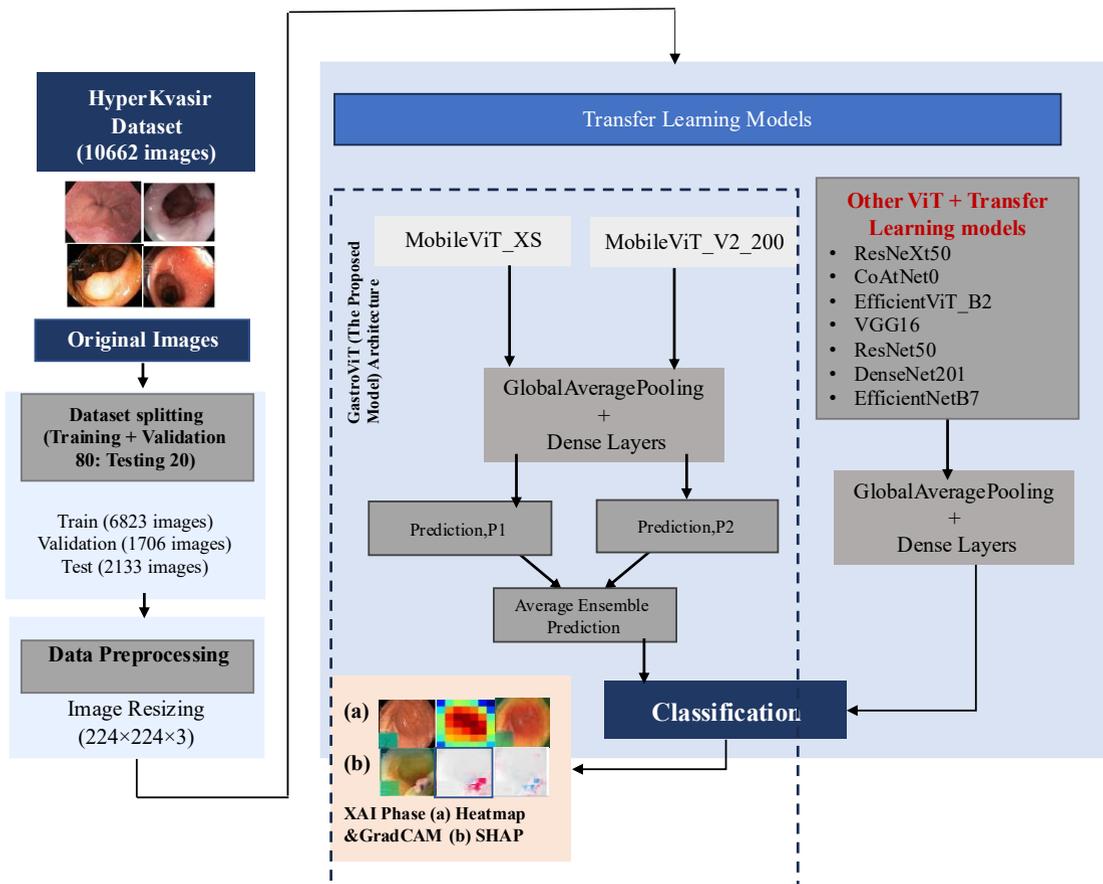

Figure 1. Proposed working framework for multi-class classification of GI diseases

MViT_XS and MViTv2_200 were used for further feature extraction while modified dense layers in the ViT models were used to classify GI tract diseases. MobileViT effectively combines the strengths of CNNs and ViTs to create a lightweight, efficient, and powerful model suitable for real-world applications, particularly on resource-constrained devices (such as mobile phones, edge devices, IoT, and so on). By merging the powers of CNNs and ViTs, it also showed superior performance. For a comprehensive examination, key metrics such as recall, accuracy, precision, F1 score, and AUC were computed. The predictions from the MViT_XS and MViTv2_200 models were ensembled to create the final prediction of the GastroViT model at the classification stage. The average ensemble method is a straightforward yet effective ensembling technique that combines predictions from numerous models by calculating the average (mean) of their outputs. Averaging can cancel out individual model errors, making the final forecast more robust. The effectiveness of the ensemble model in automated GI disease classification was carefully assessed by contrasting its performance with that of the individual ViT models and TL models. Using Grad-CAM and SHAP approaches, XAI was integrated into the final step to visually convey and interpret the model's decision-making process. The detail of each step is provided in the subsections below. Model interpretability utilizing Grad-CAM and SHAP gives doctors visual and feature-level insights, allowing them to identify which regions or factors influenced the model's predictions. This transparency builds confidence, promotes informed decision-making, and helps to validate AI-assisted diagnoses in clinical practice.

## B. DATASET AND PREPROCESSING

The proposed GastroViT framework was tested with the largest currently accessible image and video dataset of the gastrointestinal system, HyperKvasir, which was developed by Borgli et al. [29]. The information was gathered during actual colonoscopy and gastroscopy examinations at Baerum Hospital in Norway, and skilled gastrointestinal endoscopists partially annotated the images. The whole dataset includes 374 videos and 110,079 images that show anatomical landmarks along with pathological and healthy conditions.

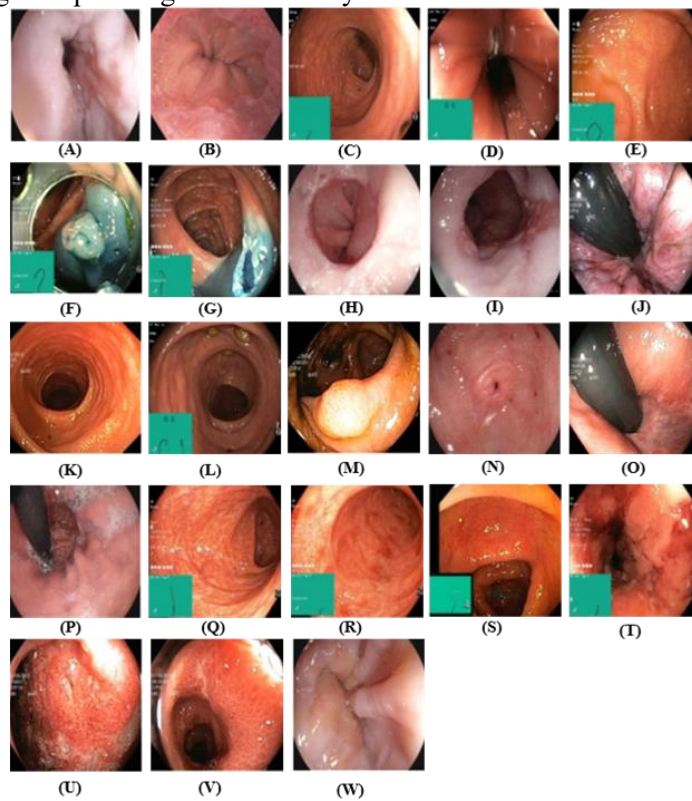

Figure 2. The HyperKvasir dataset includes (A) barretts, (B) barretts-short, (C) bbps-0-1, (D) bbps-2-3, (E) cecum, (F) dyed-lifted polyps, (G) dyed-resection margins, (H) esophagitis-a, (I) esophagitis-b-d, (J) hemorrhoids, (K) ileum, (L) impacted stool, (M) polyps, (N) pylorus, (O) retroflex-rectum, (P) retroflex stomach, (Q) ulcerative-colitis-grade-0-1, (R) ulcerative-colitis-grade-1, (S) ulcerative-colitis-grade-1-2, (T) ulcerative-colitis-grade-2, (U) ulcerative-colitis-grade-2-3, (V) ulcerative-colitis-grade-3, (W) z-line

The whole dataset is divided into four separate sections: segmented image data containing 1000 images, annotated video data containing 373 videos, labeled image data containing 10,662 images, and unlabeled image data containing 99,417 unlabeled images. In this work, only 10,662 JPEG-formatted labeled data section from the four sections were considered for classification. They are also divided into lower and upper GI tracts, characterized by pathological findings, anatomical landmarks, therapeutic interventions, and mucosal view quality, which are further labelled as 23 distinct classes, as shown in Table 1, which also addresses the total images per class. Because of some disease types occur more frequently than the others, there is a general difficulty in the medical sector in creating a balanced dataset with equal quantity of images per disease type. Figure 2 shows sample images of GI tract diseases from each class. When the samples in a particular disease class were smaller in size, the model required assistance in determining the true class label affecting the model performance. Therefore, the number of classes that had 42 training samples or fewer were removed from the dataset and a new 16 class dataset (Trial 2) were developed to compare with the model developed by the original 23 class dataset. As a result, total 6707 images of 16 classes were used for the second trial. Less training samples are frequently insufficient for robust model training, especially for deep learning models such as Vision Transformers, which require large amounts of data to master complicated features.

Table 1. The Hyperkvasir Dataset

| GI Position | Labeled Areas | Disease Types | Class Distribution |
|---|---|---|---|
| Upper GI tract | Pathological findings | barretts | 41 |
| | | barretts-short | 53 |
| | | esophagitis-a | 403 |
| | | esophagitis-b-d | 260 |
| | Anatomical landmarks | z-line | 932 |
| | | pylorus | 999 |
| | | retroflex stomach | 764 |
| Lower GI tract | Anatomical landmarks | cecum | 1009 |
| | | retroflex-rectum | 391 |
| | | ileum | 9 |
| | Pathological findings | polyps | 1028 |
| | | ulcerative-colitis-grade-0-1 | 35 |
| | | ulcerative-colitis-grade-1 | 201 |
| | | ulcerative-colitis-grade-1-2 | 11 |
| | | ulcerative-colitis-grade-2 | 443 |
| | | ulcerative-colitis-grade-2-3 | 28 |
| | | ulcerative-colitis-grade-3 | 133 |
| | | hemorrhoids | 6 |
| | Therapeutic interventions | dyed-lifted polyps | 1002 |
| | | dyed-resection margins | 989 |
| | Quality of mucosal views | bbps-0-1 | 646 |
| | | bbps-2-3 | 1148 |
| | | impacted stool | 131 |
| | | **Total** | **10,662** |

For instance, Barrell's, Barrell's-short, hemorrhoid, ileum, ulcerative colitis grades 0-1, ulcerative colitis grades 1-2, and ulcerative colitis grades 2-3 were the classes that were removed based on the criteria. Both the datasets were first split into training (80%) and testing (20%) and training sets were further split as training (80%) and validation (20%). Thus, the overall training, validation testing ratio was 64:16:20. This division ensures robust training process with sufficient data for model training and tuning, maintaining the class distribution across 23 classes. Table 2 shows overall distribution of data in the datasets with both 23 classes and 16 classes. Data augmentation was not used due to resource constraints, as low GPU availability hampered the capacity to undertake comprehensive augmentation trials.

## A. MOBILEVIT ARCHITECTURE

Mehta and Rastegari introduced MobileViT [28], which is a general-purpose, lightweight vision transformer. CNNs excel at collecting local spatial representations because they use hierarchical feature extraction via localized receptive fields. On the other hand, self-attention-based ViTs are heavyweight in nature and are employed to learn global representations. MobileViT integrates the architectures of ViTs and CNNs to create a lightweight, low-latency network that can be used for mobile vision applications by combining the advantages of CNNs with ViTs. With transformers, they provide a novel method for handling global information processing. As a result, it possesses both the efficiency and lightweight of CNNs as well as the self-attention and global vision of transformer networks, enabling it to learn both global and local features with great strength. Figure 3 presents the MobileViT architecture and modified version for this study (the way this TL model has been modified for this specific task) followed by the data preprocessing and classification block. The standard MobileViT was modified by adding Global Average Pooling (GAP) and Dense Layers for this specific task of classifying GI diseases. The MobileViT architecture, which combines ViTs and CNNs to do both local and global feature learning in a mobile-friendly manner, serves as the model that is being demonstrated in Figure 3. After capturing low-level spatial information with a 3x3 convolution layer, the network uses several MobileNetV2 (MV2) blocks for effective local feature extraction and downsampling. The MobileViT blocks are introduced at certain phases, when feature maps undergo local processing using convolutions, expand into patches, and then go through transformer layers to use multi-headed self-attention to capture global dependencies. In order to maintain spatial inductive bias, the outputs are then folded back and fused with the original features using a second convolution. The network captures rich hierarchical representations by stacking many MobileViT blocks with varying depths (L) over increasingly lower spatial resolutions. The model is accurate and efficient for mobile vision tasks thanks to the final stage, which generates the output logits using a linear classifier, global pooling, and a 1×1 convolution.

This research presents the implementation of the model MViT_XS & MViTv2_200 (which uses the separable self-attention) in transformers, for feature extraction from gastrointestinal images. Layers of fully connected, batch normalization, dropout, GAP, and softmax functions are added to complete the classification problem. For mobile vision tasks, MViT_XS, is a condensed version of the MobileViT architecture that combines the advantages of ViTs and CNNs. It makes use of the MobileViT block, which employs conventional and point-wise convolutions to initially extract local features before unfolding them into patches for transformer-based global representation learning. Multi-headed self-attention (MHA), the attention method employed here, enables each patch to attend to every other patch, thereby capturing global dependencies. Convolution is used to combine the outputs after they have been folded back. With just 2.3 million parameters, MViT_XS achieves a robust top-1 ImageNet accuracy of 74.8%, providing a good balance between accuracy and efficiency and making it ideal for devices with limited resources, such as embedded systems and smartphones.

Table 2. Overall distribution of data in the datasets with both 23 classes and 16 classes

| Testing Phase | Disease Types | Training (64%) | Validation (16%) | Testing (20%) |
|---|---|---|---|---|
| Trial 1: 23-classes | barretts | 26 | 7 | 8 |
| | barretts-short | 33 | 9 | 11 |
| | bbps-0-1 | 413 | 104 | 129 |
| | bbps-2-3 | 734 | 184 | 230 |
| | cecum | 645 | 162 | 202 |
| | dyed-lifted polyps | 641 | 161 | 200 |
| | dyed-resection margins | 632 | 159 | 198 |
| | esophagitis-a | 257 | 65 | 81 |
| | esophagitis-b-d | 166 | 42 | 52 |
| | hemorrhoids | 4 | 1 | 1 |
| | ileum | 5 | 2 | 2 |
| | impacted stool | 84 | 21 | 26 |
| | polyps | 657 | 165 | 206 |
| | pylorus | 639 | 160 | 200 |
| | retroflex-rectum | 250 | 63 | 78 |
| | retroflex stomach | 488 | 123 | 153 |
| | ulcerative-colitis-grade-0-1 | 22 | 6 | 7 |
| | ulcerative-colitis-grade-1 | 128 | 33 | 40 |
| | ulcerative-colitis-grade-1-2 | 7 | 2 | 2 |
| | ulcerative-colitis-grade-2 | 283 | 71 | 89 |
| | ulcerative-colitis-grade-2-3 | 18 | 5 | 5 |
| | ulcerative-colitis-grade-3 | 84 | 22 | 27 |
| | z-line | 596 | 150 | 186 |
| | **Total** | **6823** | **1706** | **2133** |
| Trial 2: 16-classes | bbps-0-1 | 413 | 104 | 129 |
| | bbps-2-3 | 734 | 184 | 230 |
| | cecum | 645 | 162 | 202 |
| | dyed-lifted polyps | 641 | 161 | 200 |
| | dyed-resection margins | 632 | 159 | 198 |
| | esophagitis-a | 257 | 65 | 81 |
| | esophagitis-b-d | 166 | 42 | 52 |
| | impacted stool | 84 | 21 | 26 |
| | polyps | 657 | 165 | 206 |
| | pylorus | 639 | 160 | 200 |
| | retroflex-rectum | 250 | 63 | 78 |
| | retroflex stomach | 488 | 123 | 153 |
| | ulcerative-colitis-grade-1 | 128 | 33 | 40 |
| | ulcerative-colitis-grade-2 | 283 | 71 | 89 |
| | ulcerative-colitis-grade-3 | 84 | 22 | 26 |
| | z-line | 596 | 150 | 186 |
| | **Total** | **6707** | **1676** | **2096** |

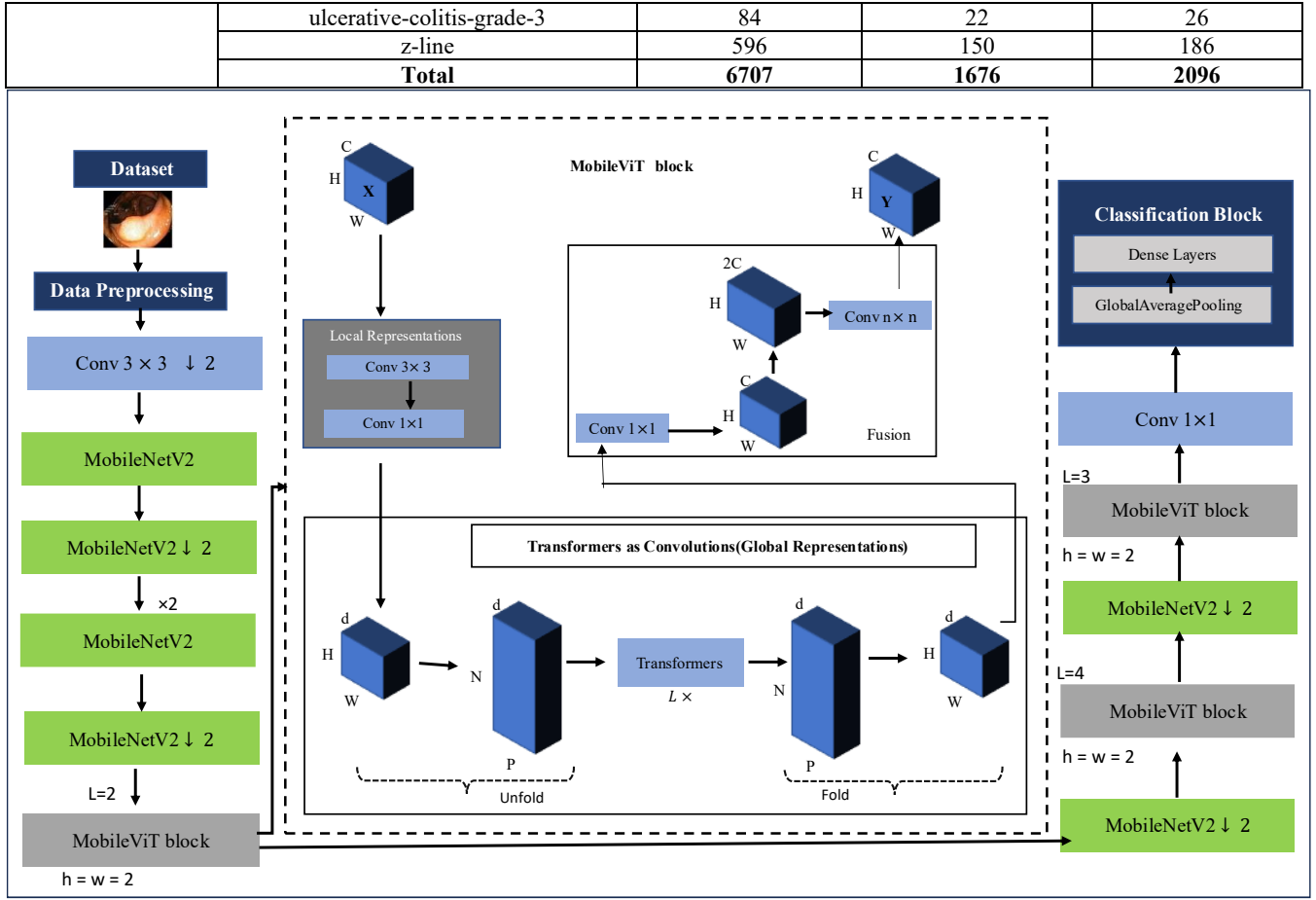

Figure 3. Modified MobileViT Architecture for proposed study

MViTv2_200 improves on the original MobileViT by substituting a separable self-attention mechanism for its multi-headed self-attention. By computing attention in relation to a latent token, separable self-attention lowers the complexity from O(k²) to O(k), where k is the number of tokens. This method is significantly faster on devices with limited resources since it does not use expensive batch-wise matrix multiplications and instead uses element-wise operations like summations and multiplications. The width multiplier that scales the model's capacity is indicated by the "200." MViTv2_200 is therefore perfect for real-time mobile AI applications.

For MViT_XS, at first local representation via convolution is done by,

$$X_{\text{local}} = \text{Conv}_{1\times1}(\text{Conv}_{n\times n}(X)) \ldots\ldots(1)$$

Here $X \in \mathbb{R}^{H\times W\times C}$ represents input feature map with height $H$, width $W$, and $C$ channels and $\text{Conv}_{n\times n}$ means a convolution capturing local spatial features (typically $3 \times 3$), $\text{Conv}_{1\times1}$ represents channel projection to a new feature dimension $C'$. $X_{\text{local}} \in \mathbb{R}^{H\times W\times C'}$ means Output contains refined local representations with fewer channels.

Then Unfolding into Patches is given by,

$$Z = \text{Unfold}(X_{\text{local}}) \in \mathbb{R}^{N\times P^2 C'} \ldots\ldots(2)$$

Here, $P \times P$ denotes the Patch size used to divide $X_{\text{local}}$ into smaller square patches and $N = \frac{H\cdot W}{P^2}$ denotes number of patches obtained after unfolding. Each patch is flattened into a vector of size $P^2 C'$, forming a matrix $Z$ of $N$ patches.

The transformer encoding is given by,

$$Z^{(l+1)} = \text{FFN}\left(\text{MSA}\left(\text{LN}(Z^{(l)})\right)\right) + Z^{(l)}, l = 1, \ldots, L \ \ldots\ldots(3)$$

where $Z^{(l)}$ means patch embeddings at the $l$-th transformer layer, LN denotes layer normalization, MSA denotes multi-head self-attention module for global interaction, FFN denotes feed-forward network applied post-attention, $L$ is the number of stacked transformer layers. Residual connections add $Z^{(l)}$ back to preserve information flow.

Then fold back into spatial grid is given by,

$$X_{\text{global}} = \text{Fold}(Z^{(L)}) \in \mathbb{R}^{H\times W\times C'} \ldots\ldots(4)$$

Where $Z^{(L)}$ means final output from transformer layers, Fold denotes rearranging the sequence $Z^{(L)}$ back into 2D spatial grid with same height and width $H \times W$, but with channel depth $C'$.
The fusion (Global + Local) is given by,
$$Y = \text{Conv}_{n \times n}(X_{\text{global}}) \ldots \ldots (5)$$

A convolution is applied on $X_{\text{global}}$ to fuse the globally attended features with learned filters. Final output $Y \in \mathbb{R}^{H \times W \times C''}$ proceeds to the next layer or block.

For MViTv2_200, Inverted Residual Block (IRB) is given by,
$$X_{\text{irb}} = X + \text{Conv}_{1 \times 1}(\text{DWConv}_{3 \times 3}(\text{Conv}_{1 \times 1}(X))) \ldots \ldots (6)$$
Where, $\text{Conv}_{1 \times 1} \to \text{DWConv}_{3 \times 3} \to \text{Conv}_{1 \times 1}$ denotes standard IRB structure with bottleneck and depthwise convolution. DWConv denotes depthwise separable convolution, reducing computation. Residual connection ensures identity mapping from input $X$ to $X_{\text{irb}}$. Linear Projections for Attention is given by,
$$Q = X_{\text{irb}} W_Q, K = X_{\text{irb}} W_K, V = X_{\text{irb}} W_V \ldots \ldots (7)$$
where, $W_Q, W_K, W_V$ denotes learned linear projection matrices to form query, key, value. Input is reshaped and projected to the token dimension $d$, resulting in: $Q, K, V \in \mathbb{R}^{N \times d}$; $N$ denotes number of tokens $= \frac{H \cdot W}{P^2}$.
Linear Attention (Efficient) is given by,
$$\text{Attention}(Q, K, V) = \phi(Q) \cdot (\phi(K)^\top V) \ldots \ldots (8)$$
where, $\phi(\cdot)$ means a kernel function (e.g., ELU, ReLU) to enable linear attention. It replaces softmax attention to reduce complexity from $O(N^2)$ to $O(N)$.
The result is attended representation using fast linear transformations.
Transformer Encoding with Residuals is given by,
$$Z^{(l+1)} = \text{FFN}\left(\text{LinearAttn}\left(\text{LN}(Z^{(l)})\right)\right) + Z^{(l)} \ldots \ldots (9)$$
It is similar to MViT_XS, but uses linear attention. It adds a feed-forward layer and skip connection.
Stack $L$ such layers to form transformer backbone.
Fusion and Output is given by,
$$Y = \text{Conv}_{1 \times 1}\left(\text{Fold}(Z^{(L)})\right) + X_{\text{irb}} \ldots \ldots (10)$$
Here, folded output reshapes token sequence $Z^{(L)}$ back to image grid. $\text{Conv}_{1 \times 1}$ denotes projecting final token embedding back to spatial channels. It adds back $X_{\text{irb}}$ (residual) to enhance gradient flow and stability.
Final Classification (Shared) is given by,
$$\hat{y} = \text{Softmax}(W_{\text{cls}} \cdot \text{GAP}(Y) + b_{\text{cls}}) \ldots \ldots (11)$$
Here, GAP $(Y)$ : Global Average Pooling across spatial dimensions. $W_{\text{cls}}, b_{\text{cls}}$ denotes classification weights and bias. It also produces class probabilities $\hat{y} \in \mathbb{R}^K$ for $K$ classes.

## C. THE PROPOSED MODEL: GASTROVIT ARCHITECTURE

The architecture of the proposed model named GastroViT, is shown in Figure 4. Initially, two pre-trained vision transformer models, were used to extract features during the feature extraction phase. These models were selected based on their capacity and being lightweight to represent complex relationships and patterns in the data. A GAP layer was then applied to the extracted features to each of the model, reducing the spatial dimensions and combining the feature maps into a single vector for each feature map. Dense layers came next, which produced class probabilities using the softmax activation function. Ultimately, the dataset was classified into 23 different illness categories by both models, guaranteeing an accurate and thorough identification of gastrointestinal ailments. By using the predictions from both models, finally, the dataset was classified into 23 different illness categories by average ensemble method, guaranteeing an accurate and thorough identification of gastrointestinal diseases. Then the average ensemble model of MViT_XS & MViTv2_200 is named by GastroViT, the proposed model.

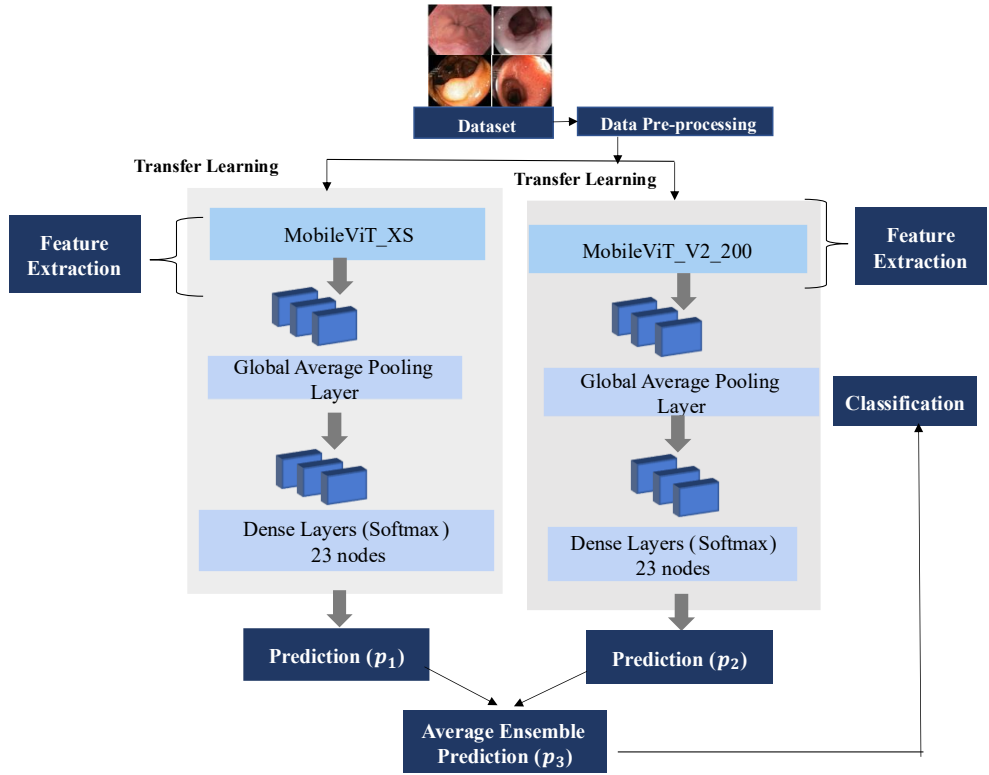

Figure 4. The proposed model architecture (GastroViT)

The benefit of the ensemble model is that it integrates the information assembled from many models. It is possible for a model to perform well for some classes and poorly for others. In ensemble learning, characteristics that are incorrectly learned by one model can still be accurately classified by combining different models and using the pattern acquired from another model. Various techniques exist for generating an ensemble model, including the following: model averaging ensemble, weighted averaging ensemble, and stacking ensemble. In our study, the average ensemble method has been adopted. The average ensemble method was used for this work because it strikes a balance between simplicity and efficiency, while alternative methods such as model averaging ensemble, weighted averaging ensemble, and stacking ensemble were not used to retain computing efficiency. The most popular and straightforward strategy is the model-averaging ensemble method. This method uses an average of the base learners' outputs to obtain the final prediction of the ensemble model. Merely averaging the ensemble models enhances the generalization performance by reducing the variation among the models, which is a result of the large variance and low bias of deep learning architectures. This is because deep learning models often overfit, resulting in lower validation accuracy and higher training accuracy. Consequently, deep learning models struggle to generalize new data. By averaging the predictions from the many base learners, this is prevented. The variance among the models is decreased by averaging the predictions, which results in accurate generalization performance. Either the predicted probabilities of the classes are averaged using the SoftMax function, or the outputs of the base learners are directly averaged. The average ensemble learning method in this investigation, which assigns each model the same weights, was used in this study. The following procedure was used to average the final softmax outputs from each model.

$$prediction = \frac{\sum P_i}{N} \quad \ldots..(12)$$

where Pi is the probability for model i and N is the total number of models. The algorithm for GastroViT is given further in below description.

### E. TRANSFER LEARNING

To improve the diagnosis of GI disorders across many classes, transfer learning models such as VGG16, DenseNet201, ResNet50, and EfficientNetB7 are essential. Due to their extensive pretraining on large datasets, these models exhibit exceptional efficiency in extracting significant characteristics from images. These models can be trained on sparse data for the intended job, which allows them to be adjusted to better capture the intricate patterns associated with GI disorders. The pretrained models were trained using more than 14 million classifications from 1,000 categories (ImageNet). There are two types of trails using transfer learning models: one by slightly modifying the models by adding additional dense layers, the other one by general TL with additional layers. To obtain accurate

**Algorithm:**

| **Algorithm for MobileViT Ensemble:** |
|---|
| Input: Dataset $\mathcal{D}$, image size $w \times h$, model set $\mathcal{M} = \{M_1, M_2\}$, train ratio $p_{\text{train}}$ |
| Output: Ensemble accuracy $\alpha_{\text{ens}}$ |
|    1. Preprocessing: |
|       1.1. Load category set $C = \{c_1, \ldots, c_k\}$. |
|       1.2. For each $c \in C$, load images $I_c$ and resize to ($w, h, 3$). |
|       1.3. Assign labels $y$ such that $y_j = \text{index}(c)$. |
|       1.4. Split $(\mathbf{X}, \mathbf{y})$ into $\mathcal{D}_{\text{train}}$ and $\mathcal{D}_{\text{test}}$ with ratio $p_{\text{train}}$. |
|    2. Training Loop: |
|       for each $M_i \in \mathcal{M}$ do |
|       2.1. Initialize optimizer AdamW($\eta, \lambda$). |
|       2.2. Compile $M_i$ with loss |
| $$L = \text{SparseCategoricalCrossentropy}$$ |
|       2.3. Train $M_i$ on $\mathcal{D}_{\text{train}}$ for $E$ epochs with validation split $p_{\text{val}}$, using callbacks: ModelCheckpoint $(W_{M_i})$, ReduceLROnPlateau |
|       2.4. Save best weights $W_{M_i}$. |
|       end for |
|    3. Prediction Loop: |
|       for each $M_i \in \mathcal{M}$ do |
|       3.1. Load weights $W_{M_i}$. |
|       3.2. Compute prediction matrix |
| $$P_i = M_i(\mathcal{D}_{\text{test}})$$ |
|       end for |
|    4. Ensemble Aggregation: |
|       4.1. Compute |
| $$P_{\text{sum}} = \sum_{i=1}^{|\mathcal{M}|} P_i$$ |
|       4.2. Obtain final class predictions |
| $$\hat{y}_{\text{ens}} = \arg\max(P_{\text{sum}}, \text{axis} = 1)$$ |
|    5. Accuracy Calculation: |
|       5.1. For each $M_i$, compute accuracy |
| $$\alpha_i = \frac{\sum_j \mathbf{1}[\hat{y}_{i,j} = y_j]}{N_{\text{test}}}$$ |
|       5.2. Compute ensemble accuracy |
| $$\alpha_{\text{ens}} = \frac{\sum_j \mathbf{1}[\hat{y}_{\text{ens},j} = y_j]}{N_{\text{test}}}$$ |
|    6. Return: $\alpha_1, \alpha_2, \alpha_{\text{ens}}$ |

classification results for the modified transfer learning model, we first integrated several dense layers of 1024 nodes, then a dropout of 0.3, and then additional dense layers equal to the number of classes during the training of the TL models which is presented in Figure 5. Transfer learning models such as VGG16, DenseNet201, ResNet50, and EfficientNetB7 were trained individually with additional dense layers to compare the proposed model with modified TL models presented at Figure 5. To measure the performance by DL models and vision transformer-based TL models such as ResNeXt50, CoAtNet0, EfficientViT_B2 following by the GAP layer & dense layers were also used individually to classify the gastrointestinal diseases. The proposed model, GastroViT, was compared to TL techniques in terms of classification results and processing resources. The performance, model parameters, layer sizes, and training and testing times are all included in this comparison. The general TL models with an additional layer demonstration is shown in Figure 6. VGG16 [31] is a deep convolutional neural network with 16 layers and is known for its simplicity and uniform architecture. It uses small 3×3 filters throughout the network, which are stacked to increase the depth and improve performance. It is commonly used for image classification tasks and as a backbone for other vision tasks.

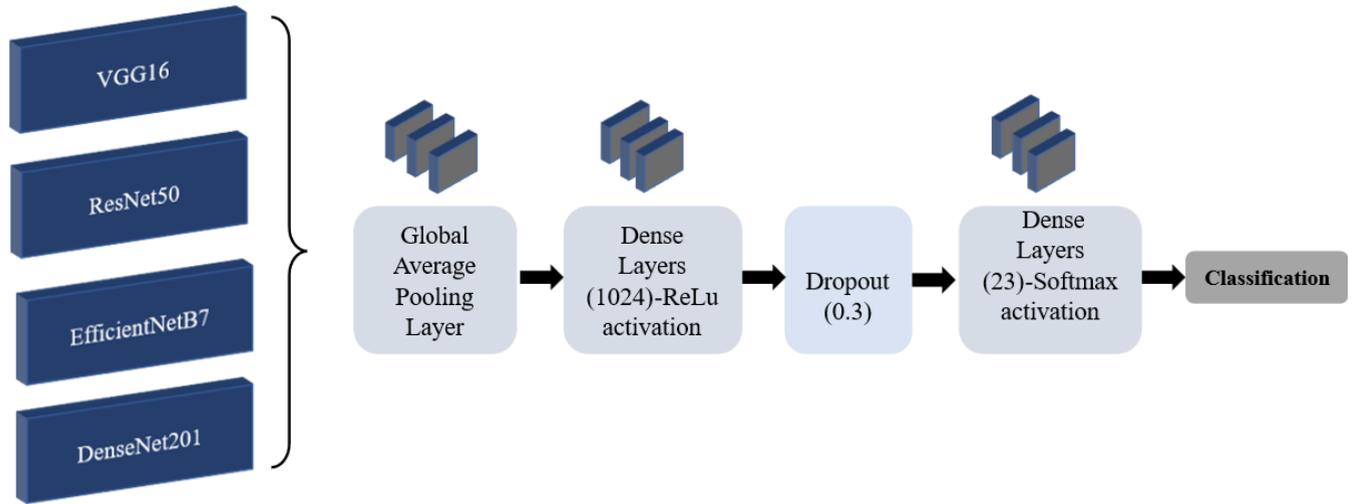

Figure 5. The modified transfer learning architecture for classifying GI diseases.

DenseNet201[31] is a densely connected convolutional network with 201 layers, where each layer is connected to every other layer in a feed-forward fashion.This dense connectivity alleviates the vanishing gradient problem, improves feature propagation, and reduces the number of parameters. It is effective for image classification, segmentation, and other computer vision tasks. ResNet50[32] is a deep residual network with 50 layers that introduces shortcut connections to bypass one or more layers. These shortcut connections help in addressing the vanishing gradient problem, allowing for very deep networks to be trained.Widely used for image classification, object detection, and other vision tasks. EfficientNetB7[33] is part of the EfficientNet family, which scales up models in a balanced manner using a compound coefficient. It achieves state-of-the-art performance by optimizing the network's width, depth, and resolution in a systematic way. It is known for its high accuracy on image classification benchmarks while being computationally efficient.ResNeXt50[34] is a convolutional neural network that enhances performance through aggregated transformations, pretrained for vision tasks.CoAtNet0[35] is a hybrid model combining convolutional and attention mechanisms for efficient and scalable vision processing, pretrained for image classification. EfficientViT_B2[36] is a vision transformer optimized for efficiency and performance, pretrained to handle high-dimensional image data with fewer parameters.

## F. IMPLEMENTATION DETAILS
### 1) HYPERPARAMETER SETTING

During the experimental study, a trial-and-error method was used to determine the ideal hyperparameters. Both the suggested and the transfer learning models were trained in identical circumstances after the optimal parameters had been determined. A thorough summary of the splitting ratios and associated accuracies for the GastroViT model can be found in Table 3. This investigation shows how the model's ability to classify gastrointestinal disorders is impacted by various data partitioning strategies. Firstly, the ratio 80:10:10 indicates that 10% was used for testing, 10% for validation, and 80% of the dataset was utilized for training. The model's accuracy was 89.68% with a split of 70:20:10. Accuracy was marginally reduced to 89.50% when validation was increased to 20% (70:20:10), indicating a small effect on training efficiency but trustworthy validation. Accuracy reduced to 88.32% with an 80:20 split (no validation set), emphasizing the necessity of validation. With a 90.57% accuracy rate, the 64:16:20 split provided the optimal ratio of testing, validation, and training.The model is rather sensitive to data partitioning, as evidenced by the moderate variation in reported accuracy across various train-validation-test splits, which ranges from 88.32% to 90.57%. Interestingly, splits with a bigger test or validation set (64:16:20, for example) produce somewhat higher accuracy, perhaps as a result of more trustworthy evaluation. These impacts aren't covered in great detail in the main text, though. In low-sample classes like "haemorrhoids" and "ileum," where there are fewer training instances, performance declines are more noticeable and the model's capacity to generalize is hampered. Data augmentation, resampling methods (oversampling minority classes or undersampling majority classes), or stratified splitting to maintain class distribution between splits are some solutions that could be used to address issue. These strategies would boost minority class performance and increase robustness against split variants.

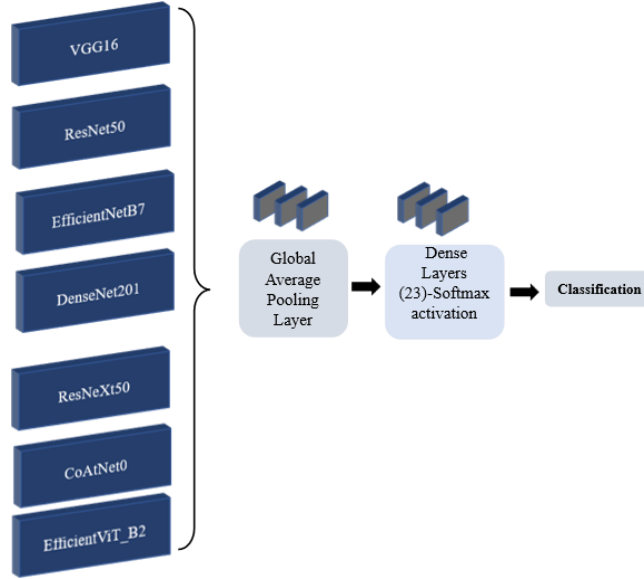

Figure 6: Transfer Learning models for classifying GI diseases.

Table 3. Splitting ratio and accuracy setting for GastroViT

| Splitting Ratio | Accuracy |
|---|---|
| 80:10:10 (Training:Validation:Testing) | 89.68% |
| 70:20:10 (Training,Validation:Testing) | 89.50% |
| 80:20 (Training:Testing) | 88.32% |
| **64:16:20(Training:Validation:Testing)** | **90.57%** |

The performance of MobileViT is compared among optimizer–loss function combinations in Table 4. The optimal configuration, AdamW with Sparse Categorical Crossentropy, outperformed other pairings with 90.57% accuracy, 63% precision, 62% recall, and 62% F1 score. A constant learning rate was applied in every experiment.

Table 4. Experiment On Different Loss Functions and Optimizers

| Optimizer | Loss Function | Precision | Recall | F1 Score | Accuracy |
|---|---|---|---|---|---|
| Adam | Sparse Categorical Crossentropy | 61% | 61% | 61% | 90.27% |
| Adam | Categorical Crossentropy | 59% | 59% | 59% | 89.59% |
| AdamW | Categorical Crossentropy | 60% | 58% | 58% | 89.26% |
| **AdamW** | **Sparse Categorical Crossentropy** | **63%** | **62%** | **62%** | **90.57%** |

The effect of different learning rates on the MobileViT model's performance parameters, such as accuracy, recall, F1 score, and precision, is summarized in Table 5 below, keeping the loss function and optimizer fixed. This table demonstrates that a learning rate of 0.001 is optimal for the proposed model.

Table 5. Experiment on Different Learning Rates

| Learning Rate | Precision | Recall | F1 Score | Accuracy |
|---|---|---|---|---|
| 0.01 | 55% | 54% | 54% | 85% |
| **0.001** | **63%** | **62%** | **62%** | **90.57%** |
| 0.0001 | 61% | 60% | 59% | 89.40% |

Hyperparameter tweaking for MobileViT is described in Table 6. Verbose logging (1) and callbacks (ModelCheckpoint, ReduceLROnPlateau with factor 0.2, patience 5, min_delta 0.0001) were employed in conjunction with ImageNet-pretrained weights. Sparse Categorical Crossentropy loss, Softmax activation, and AdamW (lr 0.001, weight decay 0.0001) were used. Following GAP, both MViT_XS (batch 64) and MViTv2_200 (batch 32) had a 23-unit dense layer; however, batch 64 was unable to run MViTv2_200 because of resource constraints of gpu. To improve performance, predictions were aggregated using an average ensemble.

Table 6. Hyperparameter Setting for GastroViT

| Parameter Name | Attribute |
|---|---|
| Weights | ImageNet |
| Verbose | 1 |
| Callbacks | ModelCheckpoint.ReduceLROnPlateau |
| Mode | max |
| Factor | 0.2 |
| Patience | 5 |
| Min Delta | 0.0001 |
| Activation Function | Softmax |
| Batch Size | 32,64 |
| Loss Function | Sparse Categorical Crossentropy |
| Optimizer | AdamW |
| Weight Decay | 0.0001 |
| Learning rate | 0.001 |

2) EXPERIMENTAL SETUP

The experiment was conducted on Kaggle's free platform, utilizing two NVIDIA T4 GPUs, each with 16 GB VRAM. This dual-GPU setup, based on the Turing architecture with tensor cores, enabled faster and more efficient deep learning training and inference compared to a single GPU, while the 8-core CPU and 16 GB RAM supported data preprocessing and parallel processing.

The GastroViT model shows moderate sensitivity to data partitioning, with accuracy ranging from 88.32% to 90.57%, and achieves optimal performance (90.57% accuracy, 63% precision, 62% recall, 62% F1) using AdamW with sparse categorical cross-entropy at a learning rate of 0.001. Experiments on Kaggle's free GPU environment demonstrate that balanced splits, effective hyperparameter tuning, and ensemble averaging significantly enhance classification robustness, especially when addressing low-sample classes.

IV. Experimental Results and Analysis

In terms of computational cost, size, layers, parameters, and accuracy, ensemble performance was compared with individual MViT_XS and MViTv2_200 models and SOTA-TL models using the GastroViT test set. Interpretability was evaluated by XAI visualizations, and comparison analysis demonstrated flexibility. Due to rare categories and a significant class imbalance, performance declined for 23-class classification, which limited generalization for rare disorders. A trade-off between accuracy and fine-grained clinical categorization was highlighted by the reduction to 16 classes, which improved all important metrics by producing a more balanced dataset but decreased diagnostic granularity.

A. EVALUATION OF THE PROPOSED MODEL GASTROVIT-TRIAL 1(23 CLASSES)

A dataset including 2,133 test images was used to evaluate each model separately. The performance of MViT_XS and MViTv2_200 was assessed through class-specific performance as shown in Table 7. The average test accuracy, precision, recall, f1-score, and AUC for the MViT_XS model were 90.57%, 63%, 62%, 62% and 94%. respectively. On the other hand, the average test accuracy, precision, recall, f1-score and AUC for the MViTv2_200 model were 90.48%, 64%, 60%, 60% and 96% respectively. Table 7 shows a significant difference in categorization performance between classes. Classes with high precision, recall, and F1 scores—such as pylorus, polyps, and bbps-0-1—are easier to classify. Classes like barretts, barretts-short, haemorrhoids, and several ulcerative colitis grades, on the other hand, have extremely low or zero ratings, indicating serious difficulties that are probably brought on by class imbalance, minute feature variations, or unclear labelling. The ensemble model, which achieves the best overall accuracy of 91.98%, typically improves results by integrating the capabilities of multiple models. This implies that rare or visually identical classes need more work to increase classification accuracy, even while the models perform well on distinct and well-represented classes. Figure 7 shows the training and validation accuracy curves for MViT_XS and MViTv2_200 for 23 classes. Figure 8 and Figure 9 show the class-wise confusion matrix and ROC curves for MViT_XS and MViTv2_200. The proposed ensemble model GrastroViT, achieved a precision, recall, and F1 score of 69%, 63%, and 64%, respectively, and an accuracy of 91.98%, as presented in Table 7 for twenty-three classes. The proposed ensemble model GrastroViT has a higher accuracy than MViTv2_200 & MViT_XS by 0.41%, 0.50%.

Table 7. Performance of the twenty-three classes of MobileViT_XS and MobileViT_V2_200

| Class Name | MobileViT_XS | | | MobileViT_V2_200 | | | Ensemble of MobileViT_XS & MobileViT_V2_200 | | |
|---|---|---|---|---|---|---|---|---|---|
| | Precision | Recall | F1 Score | Precision | Recall | F1 Score | Precision | Recall | F1 Score |
| barretts | 0.33 | 0.25 | 0.29 | 0.50 | 0.12 | 0.20 | 0.50 | 0.25 | 0.33 |
| barretts-short | 0.25 | 0.18 | 0.21 | 0.25 | 0.09 | 0.13 | 1.00 | 0.27 | 0.43 |
| bbps-0-1 | 0.98 | 0.98 | 0.98 | 1.00 | 0.97 | 0.94 | 0.98 | 0.98 | 0.98 |
| bbps-2-3 | 0.98 | 0.98 | 0.98 | 0.96 | 0.98 | 0.97 | 0.97 | 0.98 | 0.98 |
| cecum | 0.98 | 0.98 | 0.98 | 0.97 | 1.00 | 0.98 | 0.99 | 0.98 | 0.98 |
| dyed-lifted polyps | 0.91 | 0.96 | 0.94 | 0.95 | 0.94 | 0.95 | 0.95 | 0.97 | 0.96 |
| dyed-resection margins | 0.96 | 0.91 | 0.94 | 0.95 | 0.95 | 0.95 | 0.97 | 0.94 | 0.96 |
| esophagitis-a | 0.59 | 0.58 | 0.58 | 0.65 | 0.35 | 0.45 | 0.68 | 0.60 | 0.64 |
| esophagitis-b-d | 0.75 | 0.58 | 0.65 | 0.61 | 0.73 | 0.67 | 0.76 | 0.62 | 0.68 |
| hemorrhoids | 0.00 | 0.00 | 0.00 | 0.00 | 0.00 | 0.00 | 0.00 | 0.00 | 0.00 |
| ileum | 0.00 | 0.00 | 0.00 | 0.00 | 0.00 | 0.00 | 0.00 | 0.00 | 0.00 |
| impacted stool | 0.81 | 0.85 | 0.83 | 0.79 | 0.73 | 0.76 | 0.81 | 0.81 | 0.81 |
| polyps | 0.99 | 0.98 | 0.98 | 1.00 | 0.98 | 0.99 | 1.00 | 0.98 | 0.99 |
| pylorus | 0.99 | 0.99 | 0.99 | 0.96 | 1.00 | 0.98 | 1.00 | 1.00 | 1.00 |
| retroflex-rectum | 0.94 | 0.99 | 0.96 | 0.96 | 1.00 | 0.98 | 0.96 | 1.00 | 0.98 |
| retroflex stomach | 0.98 | 0.99 | 0.98 | 0.99 | 0.97 | 0.98 | 0.99 | 0.97 | 0.98 |
| ulcerative-colitis-grade-0-1 | 0.33 | 0.14 | 0.20 | 0.33 | 0.14 | 0.20 | 0.50 | 0.14 | 0.22 |
| ulcerative-colitis-grade-1 | 0.45 | 0.57 | 0.51 | 0.68 | 0.38 | 0.48 | 0.58 | 0.62 | 0.60 |
| ulcerative-colitis-grade-1-2 | 0.00 | 0.00 | 0.00 | 0.00 | 0.00 | 0.00 | 0.00 | 0.00 | 0.00 |
| ulcerative-colitis-grade-2 | 0.74 | 0.71 | 0.72 | 0.64 | 0.93 | 0.76 | 0.65 | 0.82 | 0.73 |
| ulcerative-colitis-grade-2-3 | 0.00 | 0.00 | 0.00 | 0.00 | 0.00 | 0.00 | 0.00 | 0.00 | 0.00 |
| ulcerative-colitis-grade-3 | 0.67 | 0.74 | 0.70 | 0.82 | 0.52 | 0.64 | 0.74 | 0.52 | 0.61 |
| z-line | 0.85 | 0.93 | 0.89 | 0.78 | 0.94 | 0.85 | 0.84 | 0.98 | 0.91 |
| **Average±SD** | **0.63±0.35** | **0.62±0.39** | **0.62±0.38** | **0.64±0.37** | **0.60±0.42** | **0.60±0.39** | **0.69±0.36** | **0.63±0.39** | **0.64±0.38** |
| AUC | | 0.94 | | | 0.96 | | | - | |
| **Test Accuracy(%)** | | **90.57** | | | **90.48** | | | **91.98** | |

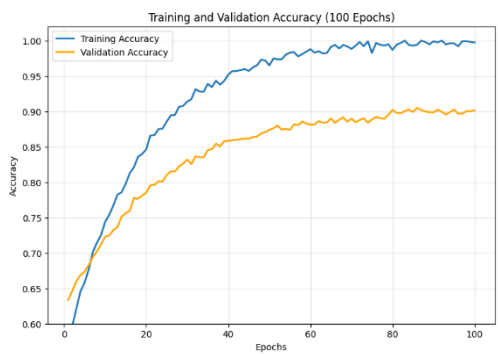

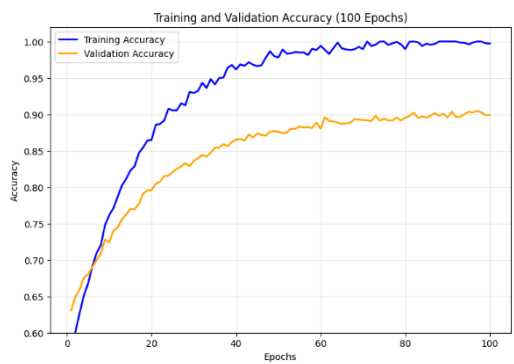

(a)  (b)

Figure 7: Training and validation accuracy curve for 23 classes a) MobileViT_XS b) MobileViT_V2_200

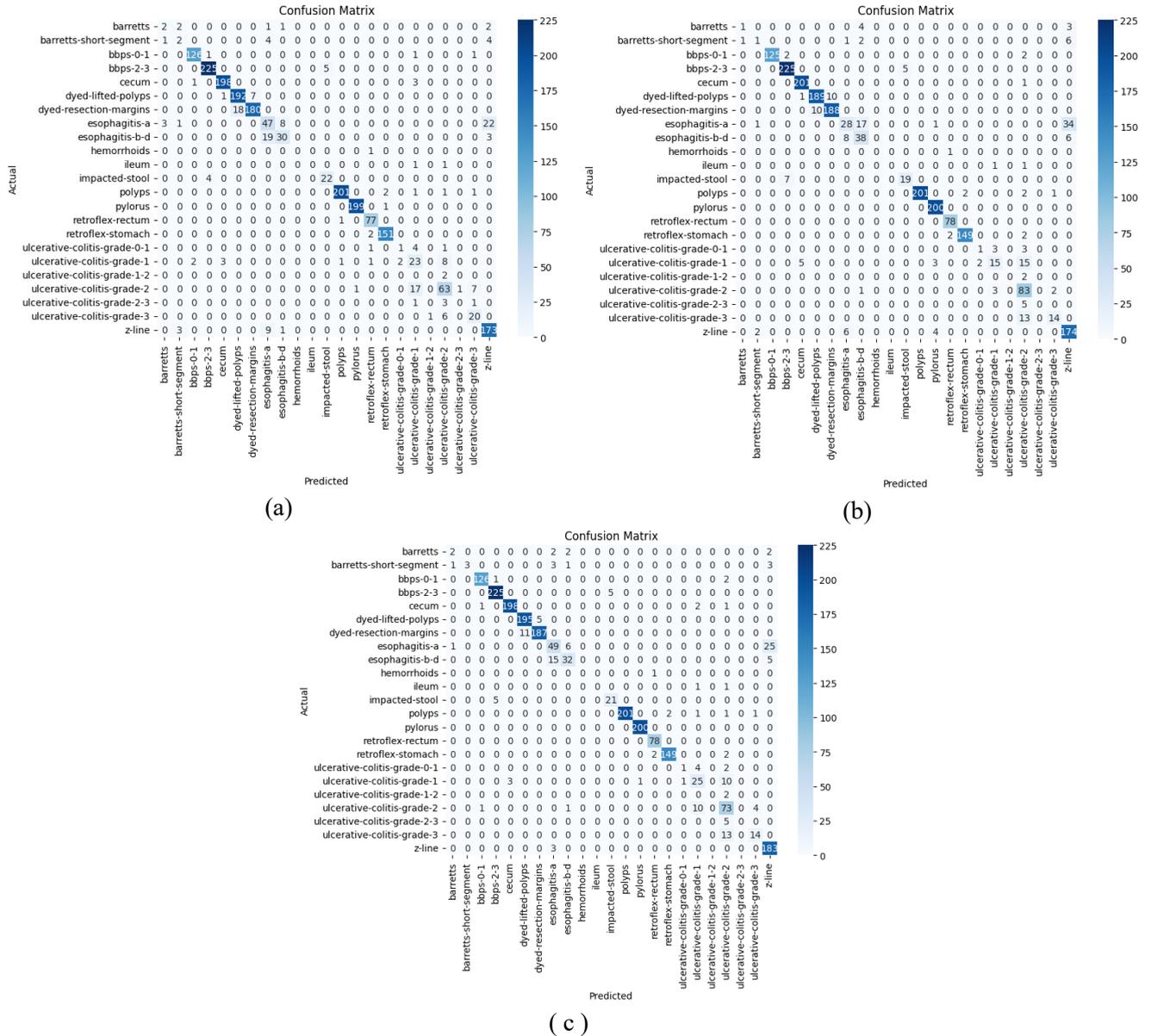

Figure 8. Confusion matrix of models for twenty-three classes. (a) MobileViT_XS (b) MobileViT_V2_200 (c) Proposed ensemble model

The overall performance of several ensemble approaches on MViT_XS and MViTv2_200 is compared in Table 8. The average ensemble outperforms the weighted average and stacked ensembles, with the maximum test accuracy of 91.98%, the best precision (0.69), and the best F1 score (0.64). This implies that more balanced and useful predictions for this job are produced by a simple average ensemble.

Table 8. Overall performance comparison between various ensemble

| Model Name | Precision | Recall | F1 Score | Testing accuracy |
|---|---|---|---|---|
| Weighted average ensemble | 0.62±0.3646 | 0.61±0.3833 | 0.61±0.3880 | 89.50% |
| Stacked ensemble | 0.63±0.3476 | 0.60±0.4055 | 0.59±0.320 | 88.60% |
| **Average ensemble** | **0.69±0.3620** | **0.63±0.3929** | **0.64±0.3756** | **91.98%** |

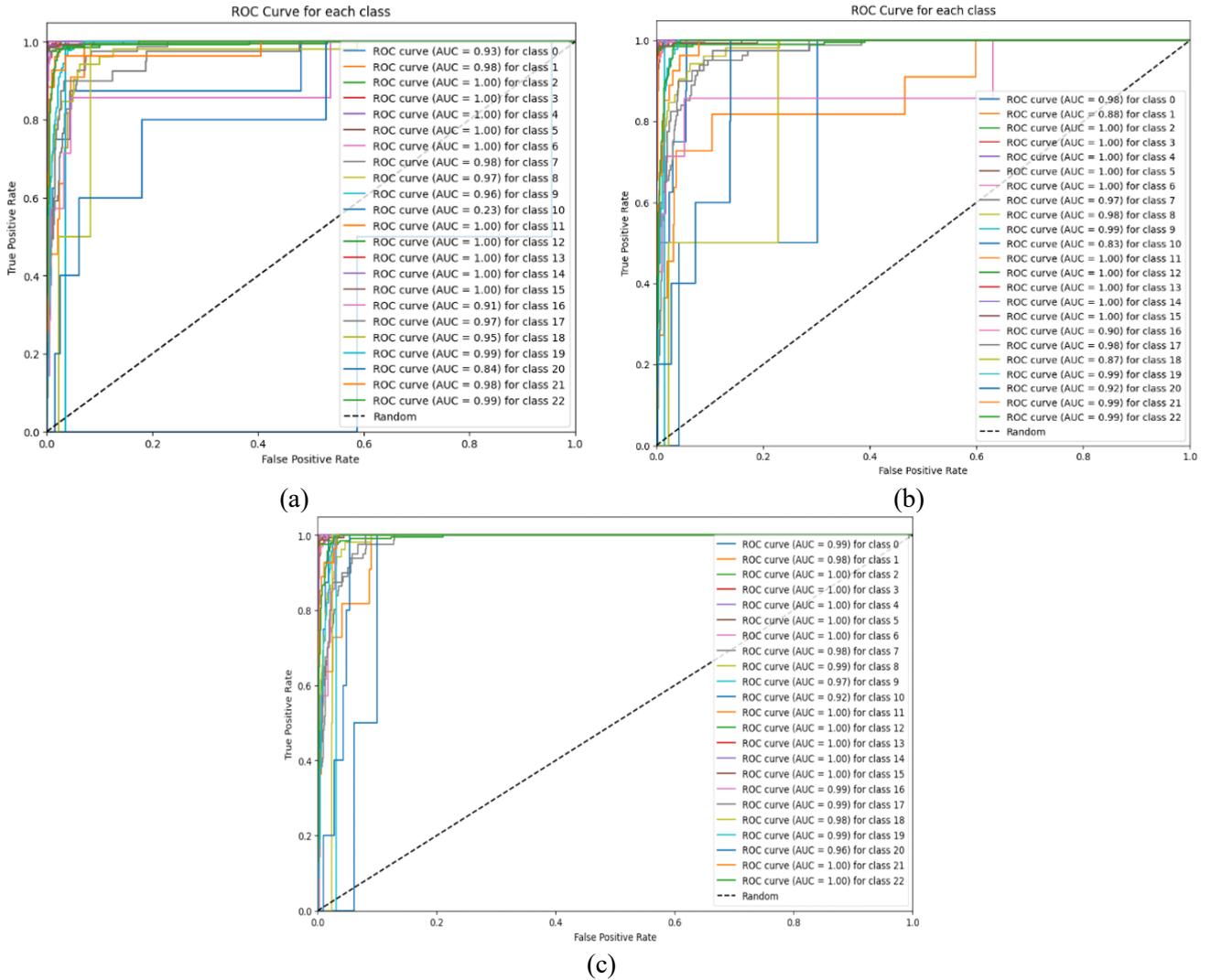

Figure 9. ROC curves of the models for the twenty-three classes. (a) MobileViT_XS (b) MobileViT_V2_200 (c)Proposed ensemble model

## B. EVALUATION OF THE PROPOSED MODEL GASTROVIT-TRIAL 1(16 CLASSES)

MViT_XS & MViTv2_200 models were trained on a dataset of 6,707 images that corresponded to 16 distinct disease categories. A dataset including 2,096 test images was used to evaluate each model separately. The performance of MViT_XS and MViTv2_200 was assessed through a class-specific approach as shown in Table 9 for 16 classes.

The average precision,recall and F1 score for the MViT_XS model were 86%, 85%, and 86%, respectively presented in Table 9. The AUC was 99%, while the accuracy was 92.23%. On the other hand, the average precision, recall, and f1-score for the MViTv2_200 model were 86%, 87%, and 86%, respectively. The AUC was 99%, while the accuracy was 92.12%. Figure 10 and Figure 11 show the confusion matrix and ROC curves for each class for MViT_XS and MViTv2_200 for 16 classes. The proposed ensemble model GastroViT achieved a precision, recall, and F1 score of 87%, 86%, and 87%, respectively, and an accuracy of 92.70%, as presented in Table 9 for sixteen classes. Due to class imbalance and the elimination of low-sample classes, the ensemble model in Table 9 only slightly improves accuracy when compared to the individual MobileViT models. There is little opportunity for ensemble gains because the dataset is dominated by well-represented classes where both models already exhibit good performance. The potential value of the ensemble is diminished when under-represented groups perform worse and contribute less to total accuracy. More direct attention to class disparity may help bring attention to the real benefits of ensembling.The class imbalance is the main cause of the lower precision, recall, and F1 score in the first experiment with 23 classes; the model had trouble properly identifying some diseases classes when there were limited data, which resulted in misclassifications and poorer performance overall.

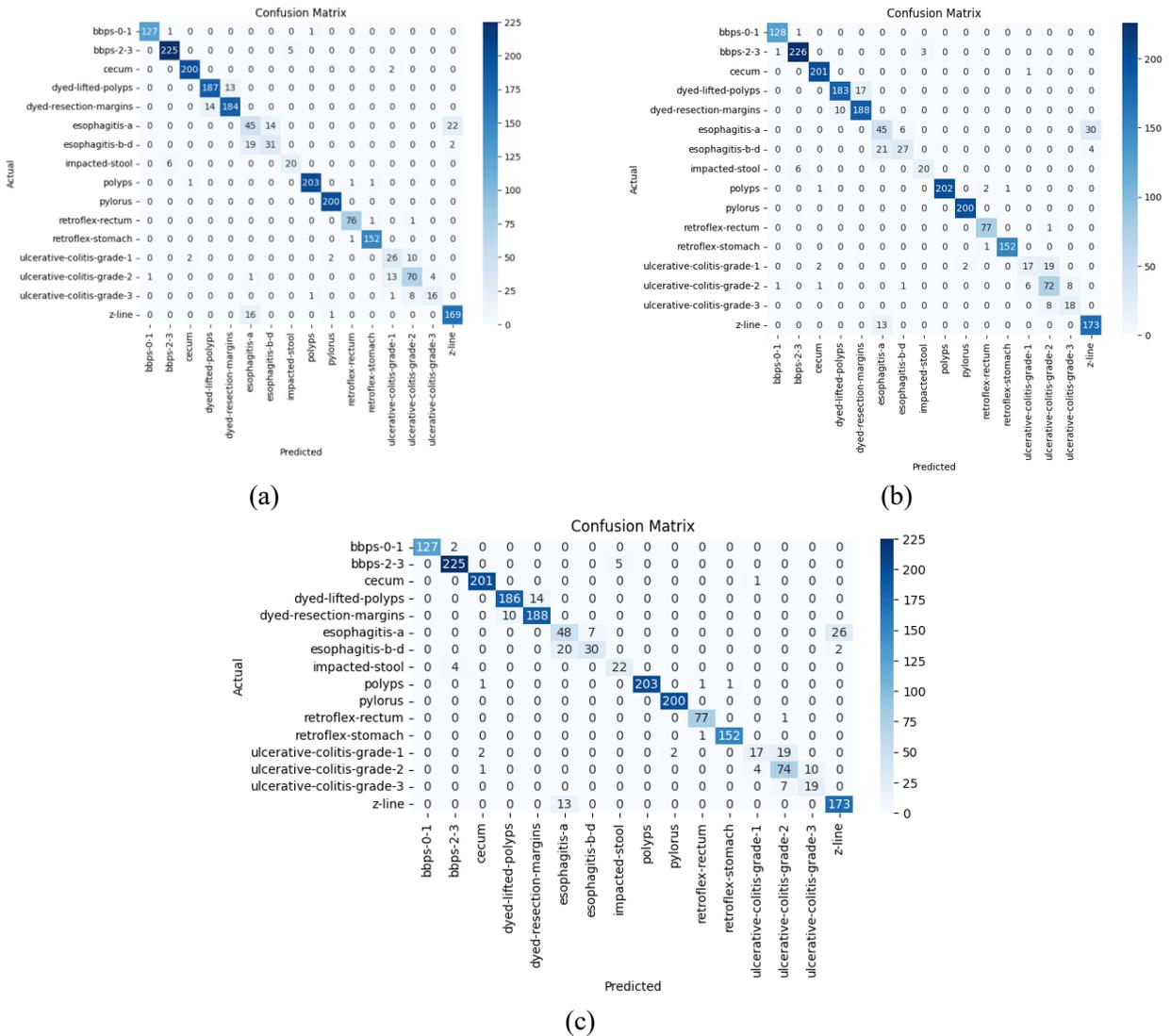

Figure 10. Confusion matrix of the models for sixteen classes. (a) MobileViT_XS (b) MobileViT_V2_200 (c) proposed ensemble model

## C. COMPARISON WITH TRANSFER LEARNING MODELS AND GASTROVIT

Table 10, 11(See appendix) presents a comparison of modified CNN-based pretrained models and CNN based TL models across 23 classes. Table 12(See appendix) presents a comparison of ViT-based pretrained models where EfficientViT_B2 achieves the highest value. Some significant differences that illustrate the models' advantages and disadvantages in various settings may be seen when comparing the four models— the modified VGG16, DenseNet201, EfficientNetB7, and ResNet50—across a range of performance indicators from Table 10. DenseNet201 beats the other models in terms of test accuracy, scoring 88.16%, then ResNet50 (87.50%), EfficientNetB7 (87.02%), and VGG16 (86.55%) in Table 10 for modified TL. Table 11 presents a performance comparison of four deep learning models (VGG16, DenseNet201, EfficientNetB7, and ResNet50) for GI disease classification without modification. It includes precision, recall, and F1-score for each disease class, along with average performance metrics, AUC (Area Under Curve), and test accuracy. Among the models, ResNet50 achieves the highest test accuracy (86.60%), followed by EfficientNetB7 (85.51%), DenseNet201 (84.89%), and VGG16 (83.19%). The comparison of ResNeXt50, EfficientViT_B2, and CoAtNet0 models reveals distinct strengths and weaknesses across various medical imaging classification tasks presented in Table 12. EfficientViT_B2 showing a slight edge in testing accuracy at 89.02%, compared to ResNeXt50's 88.79%. CoAtNet0, while trailing in F1 score, achieves an accuracy of 81.85%, indicating a more pronounced drop in performance across the board.

The comparative analysis of various models, including VGG16, ResNet50, EfficientNetB7, DenseNet201, CoAtNet0, EfficientViT_B2, ResNeXt50, MViT_XS, MViTv2_200, and MobileViT_ensemble, highlights differences in precision, recall, F1 score, and testing accuracy in Table 13.MobileViT_ensemble stands out as the top performer, achieving the highest precision (0.69), recall (0.63), F1 score (0.64), and accuracy (91.98%), demonstrating the effectiveness of model ensembling. MViTv2_200 and MViT_XS also show strong performance, with F1 scores of 0.60 and 0.62, respectively, and testing accuracy exceeding 90%. Vision Transformers surpass CNNs by capturing global contextual information, which is useful for detecting complex patterns throughout a picture. However, they may struggle with little data and nuanced local features, where CNNs' strong inductive biases and local feature extraction are advantageous

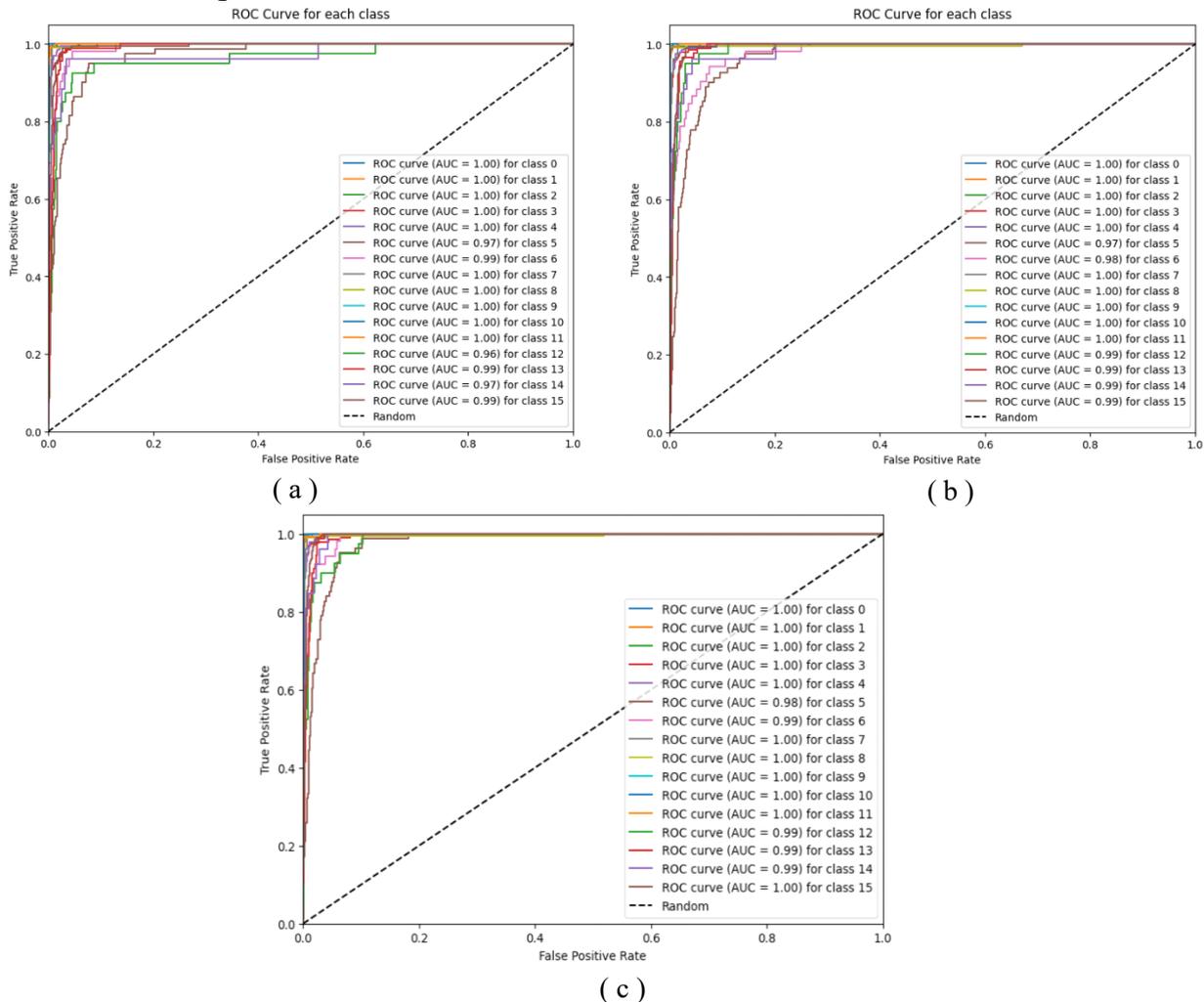

Figure 11. ROC curves of the models for sixteen classes. (a) MobileViT_XS (b) MobileViT_V2_200 (c) MobileViT_ensemble

CoAtNet0 also shows weaker performance, with the lowest F1 score (0.52) and testing accuracy (81.85%). Models like ResNet50, EfficientNetB7, and DenseNet201 offer a middle ground, with F1 scores ranging from 0.54 to 0.58 and testing accuracies around 85-88%. EfficientViT_B2 and ResNeXt50 show comparable performance, with EfficientViT_B2 slightly outperforming in testing accuracy at 89.02%. The main reason MViT_XS and MViTv2_200 were chosen as baselines was because of their small model sizes, which make them ideal for light-weight and effective vision tasks.While MViTv2_200 integrates self-separable attention for improved feature extraction at a lower computational cost, MViT_XS places more emphasis on self-attention techniques to efficiently capture long-range relationships. All things considered, the MobileViT architecture blends local feature learning with global context modelling, deftly combining the advantages of both CNNs and Vision Transformers. These models are positioned as robust, resource-efficient baselines for assessing GastroViT in the context of the current lightweight Transformer landscape because of this balance.While lightweight models like MViT_XS and MViTv2_200 offer reduced computational costs, their limited capacity can lead to lower F1 scores, especially for rare classes. This highlights a trade-off between model efficiency and the ability to accurately learn features from small sample categories.

Table 9. Sixteen-class performance by MobileViT_XS, MobileViT_V2_200, MobileViT_ensemble

| Class Name | MobileViT_XS | | | MobileViT_V2_200 | | | Ensemble of MobileViT_XS & MobileViT_V2_200 | | |
|---|---|---|---|---|---|---|---|---|---|
| | Precision | Recall | F1 Score | Precision | Recall | F1 Score | Precision | Recall | F1 Score |
| bbps-0-1 | 1.00 | 0.99 | 1.00 | 0.97 | 0.99 | 0.98 | 0.98 | 0.99 | 0.98 |
| bbps-2-3 | 0.96 | 0.99 | 0.98 | 0.98 | 0.97 | 0.97 | 0.98 | 0.97 | 0.97 |
| cecum | 0.98 | 0.99 | 0.98 | 0.98 | 0.99 | 0.98 | 0.98 | 0.99 | 0.98 |
| dyed-lifted polyps | 0.97 | 0.90 | 0.93 | 0.94 | 0.94 | 0.94 | 0.94 | 0.94 | 0.94 |
| dyed-resection margins | 0.90 | 0.97 | 0.93 | 0.94 | 0.94 | 0.94 | 0.94 | 0.94 | 0.94 |
| esophagitis-a | 0.65 | 0.52 | 0.58 | 0.64 | 0.48 | 0.55 | 0.64 | 0.58 | 0.61 |
| esophagitis-b-d | 0.67 | 0.65 | 0.66 | 0.74 | 0.75 | 0.74 | 0.76 | 0.62 | 0.68 |
| impacted stool | 0.86 | 0.69 | 0.77 | 0.79 | 0.88 | 0.84 | 0.76 | 0.85 | 0.80 |
| polyps | 1.00 | 0.99 | 0.99 | 1.00 | 0.98 | 0.99 | 1.00 | 0.98 | 0.99 |
| pylorus | 1.00 | 1.00 | 1.00 | 0.99 | 1.00 | 1.00 | 1.00 | 1.00 | 1.00 |
| retroflex-rectum | 0.97 | 0.99 | 0.98 | 0.95 | 0.99 | 0.97 | 0.96 | 0.99 | 0.97 |
| retroflex stomach | 0.99 | 0.99 | 0.99 | 0.99 | 0.99 | 0.99 | 0.99 | 0.99 | 0.99 |
| ulcerative-colitis-grade-1 | 0.52 | 0.57 | 0.55 | 0.57 | 0.65 | 0.60 | 0.62 | 0.67 | 0.60 |
| ulcerative-colitis-grade-2 | 0.75 | 0.71 | 0.73 | 0.81 | 0.69 | 0.74 | 0.78 | 0.76 | 0.77 |
| ulcerative-colitis-grade-3 | 0.75 | 0.69 | 0.72 | 0.69 | 0.69 | 0.69 | 0.72 | 0.69 | 0.71 |
| z-line | 0.87 | 0.95 | 0.91 | 0.83 | 0.92 | 0.87 | 0.87 | 0.95 | 0.91 |
| Average±sd | 0.86±0.1517 | 0.85±0.1760 | 0.86±0.1607 | 0.86±0.1412 | 0.87±0.1609 | 0.86±0.1507 | 0.87±0.1348 | 0.86±0.1515 | 0.87±0.1456 |
| AUC | 0.99 | | | 0.99 | | | - | | |
| Test Accuracy(%) | 92.23 | | | 92.12 | | | 92.70 | | |

Table 13. Overall performance comparison between all models for 23 classes

| Model Name | Precision | Recall | F1 Score | Testing accuracy |
|---|---|---|---|---|
| Modified VGG16 | 0.56±0.4075 | 0.57±0.4143 | 0.56±0.4094 | 86.55% |
| Modified ResNet50 | 0.58±0.4022 | 0.58±0.4153 | 0.58±0.4073 | 87.5% |
| Modified EfficientNetB7 | 0.61±0.4026 | 0.57±0.4125 | 0.57±0.4041 | 87.02% |
| Modified DenseNet201 | 0.60±0.3943 | 0.57±0.4169 | 0.58±0.4051 | 88.16% |
| VGG16 | 0.55±0.3911 | 0.54±03956 | 0.54±0.3930 | 83.19% |
| ResNet50 | 0.59±0.3747 | 0.57±0.4231 | 0.56±0.3985 | 86.60% |
| EfficientNetB7 | 0.57±0.3907 | 0.55±0.4189 | 0.55±0.4026 | 85.51% |
| DenseNet201 | 0.58±0.3990 | 0.54±0.4222 | 0.54±0.4035 | 84.89% |
| CoAtNet0 | 0.55±0.3948 | 0.52±0.3834 | 0.52±0.3790 | 81.85% |
| EfficientViT_B2 | 0.60±0.3967 | 0.59±0.4150 | 0.59±0.4063 | 89.02% |
| ResNeXt50 | 0.57±0.4080 | 0.58±0.4162 | 0.58±0.4106 | 88.79% |
| MovileViT_XS | 0.63±0.3746 | 0.62±0.3933 | 0.62±0.3840 | 90.57% |
| MobileViT_V2_200 | 0.64±0.3676 | 0.60±0.4155 | 0.60±0.3910 | 90.48% |
| **MobileViT_ensemble** | **0.69±0.3620** | **0.63±0.3929** | **0.64±0.3756** | **91.98%** |

### *D. COMPUTATIONAL TIME AND RESOURCE ALLOCATION*

Table 14 provides a comparison of various neural network models based on several metrics like model name, total parameters, trainable parameters and testing time. The total number of parameters in the model, which includes both trainable and non-trainable parameters. This is a measure of the model's complexity and capacity. The number of parameters that are updated during training. This indicates the model's ability to learn from data. The size of the model in megabytes, reflecting how much storage space the model requires. The average time it takes for the model to process a single input during inference. This is an indicator of the model's efficiency in making predictions. MobileViT_ensemble provides the best performance, but it requires more computational resources and may have longer inference times due to its ensemble structure. CoAtNet0, on the other hand, performs less precisely but is likely faster and more resource-efficient, making it appropriate for situations with limited hardware or strict real-time needs. Modified models often improve performance but may increase computing costs. Finally, the optimal option is determined by whether high accuracy or computing efficiency is the most important consideration for your particular application. Every single Vision Transformer (ViT) version and transfer learning (TL) model is trained and tested using the same experimental conditions, including dataset splits, preprocessing, and evaluation metrics, as part of the benchmarking process. This guarantees that architectural variances, not experimental bias, are the only cause of performance discrepancies. The same technique is then used to evaluate the ensemble model, allowing for a straightforward and equitable comparison with its individual counterparts.

TL models are compared by inference time and size (parameters) in Table 14. VGG16 is older and has a moderate level of complexity (15.3M parameters). ResNet50 (25.7M) is deeper, while EfficientNetB7 (66.7M) has the most trainable parameters. With CoAtNet0 testing in 0.0072s, DenseNet201 (20.3M) and CoAtNet0 (22.5M) are both effective. ResNeXt50 is a quicker version of ResNet50. Both MViT_XS and MViTv2_200 are small; MViTv2_200 tests in 0.003 seconds. The ensemble averages 0.01s each sample and has 20M parameters. Although larger versions typically have greater capacity, they also demand more processing power.

Table 14. Computational Resources and Time Comparison for Multiclass Classifications

| Model Name | Total Parameters (Million) | Trainable Parameters (Million) | Testing Time (Second per sample) |
|---|---|---|---|
| VGG16 | 15.26 | 0.55 | 0.023 |
| ResNet50 | 25.71 | 2.12 | 0.012 |
| EfficientNetB7 | 66.74 | 2.65 | 0.019 |
| DenseNet201 | 20.31 | 0.03 | 0.019 |
| CoAtNet0 | 22.54 | 22.53 | 0.0072 |
| ResNeXt50 | 23.10 | 23.03 | 0.0049 |
| EfficientViT_B2 | 21.86 | 21.83 | 0.0052 |
| MovileViT_XS | 1.95 | 1.94 | 0.007 |
| MobileViT_V2_200 | 1.48 | 17.45 | 0.003 |
| **MobileViT_ensemble** | **20** | **19.39** | **0.01** |

*E. Interpretability of GastroViT by SHAP & Grad CAM*

A Grad-CAM analysis uses gradients from the last convolution layer to identify the activations that contributed most to the prediction of a particular category [37]. In the ever-evolving landscape of deep learning and artificial intelligence, the quest for model interpretability and transparency is paramount. Grad-CAM (gradient-weighted class activation mapping) visualization has emerged as a powerful tool, offering a window into the intricate decision-making processes of deep neural networks. By highlighting the regions crucial for classification, this approach bridges the gap between model predictions and human interpretation. With Grad-CAM, the black box of deep learning models becomes slightly more transparent, shedding light on where the model focuses its attention to make decisions. On the other hand, A technique for explaining the results of machine learning models, particularly deep learning, is called SHAP (SHapley Additive exPlanations)[38]. Red pixels in SHAP visualizations denote characteristics that boost the model's output and imply a better prediction score. Features that reduce the model's output are represented by blue pixels, which denote a lower prediction score. The model's predictions for the top classes can be interpreted using the SHAP. The explanation allocates SHAP values to distinct picture regions for every class that is highly predicted, signifying their influence on the class's prediction. Based on the most significant features, the model most confidently predicts the first top classes with the highest SHAP values. Based on the features that have the most impact on the model's judgement, as shown by the SHAP values, the five most top classes prediction by the model has been ranked for each input image which is shown in Figure 12. Random samples were chosen in the SHAP explanation in order to make predictions and evaluate the interpretability of the suggested approach. Pixelated visualizations were produced after a thorough examination of multiple GI features that yielded Shapley values. An assortment of test samples was chosen at random for prediction in order to get the SHAP results. The SHAP results are shown in Figure 12 with the original images mixed with a light grey backdrop. The top five classes predictions are explained by SHAP. The presence of dyed resection margins is shown by the red pixels in the top row of the SHAP explanation images. Red pixels were shown in both classes, dyed lifted polyps and dyed resection margins showed a very competitive XAI prediction. Nonetheless, there are a lot of blue pixels in dyed lifted polyps. On the other hand, other class groupings were correctly eliminated due to the absence of blue pixels and the decrease in red pixels. The z-line class is represented by red pixels in the SHAP explanation images, as demonstrated in the third row, where an excess of red pixels accurately indicates class membership. Consequently, a lesser likelihood was indicated by blue pixels in the SHAP explanation visuals for the other classes. The fourth row of the SHAP explanation images' red pixels indicated a strong indication of dyed resection class GI disease. Additionally, by highlighting red pixels in particular regions, all rows successfully identified disease classes.

The suggested approach occasionally attends to irrelevant or deceptive regions, resulting in inaccurate predictions, as shown by the SHAP visualizations for misclassified images in Figure 13. Uncertainty in feature localization is indicated by the frequent scattering of high positive (red) and negative (blue) contributions. Confusion is exacerbated by similar visual patterns across related classes (such as barretts and barretts-short-segment). This emphasizes the need for more accurate focus on lesion-relevant areas and improved feature discrimination. The analysis showed a distinct pattern: blue pixels suggested a lesser likelihood of belonging to the target class, whereas red pixels effectively indicated specific classes.

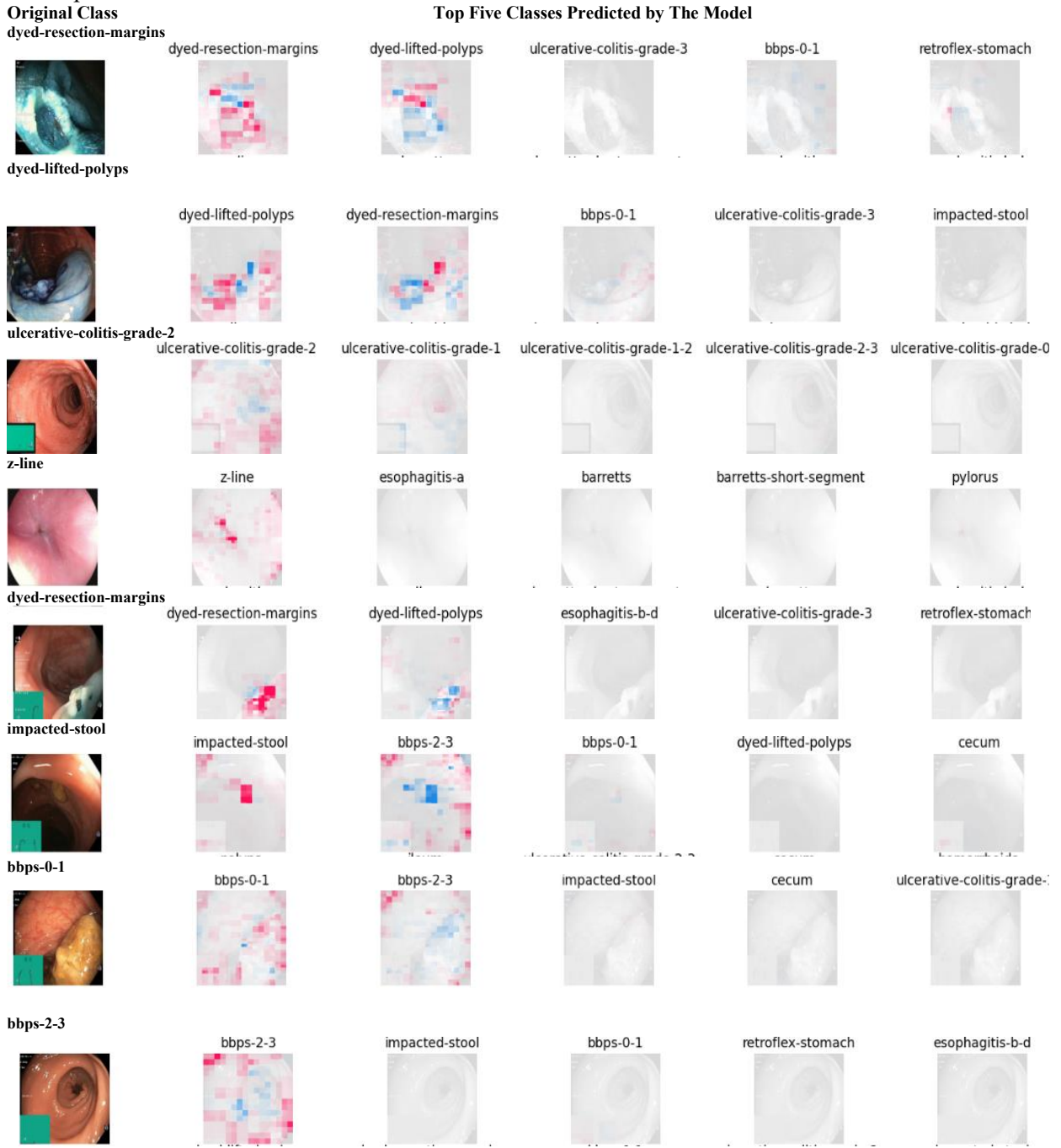

Figure 12. SHapley Additive exPlanations (SHAP) accurately predicted images for the proposed model

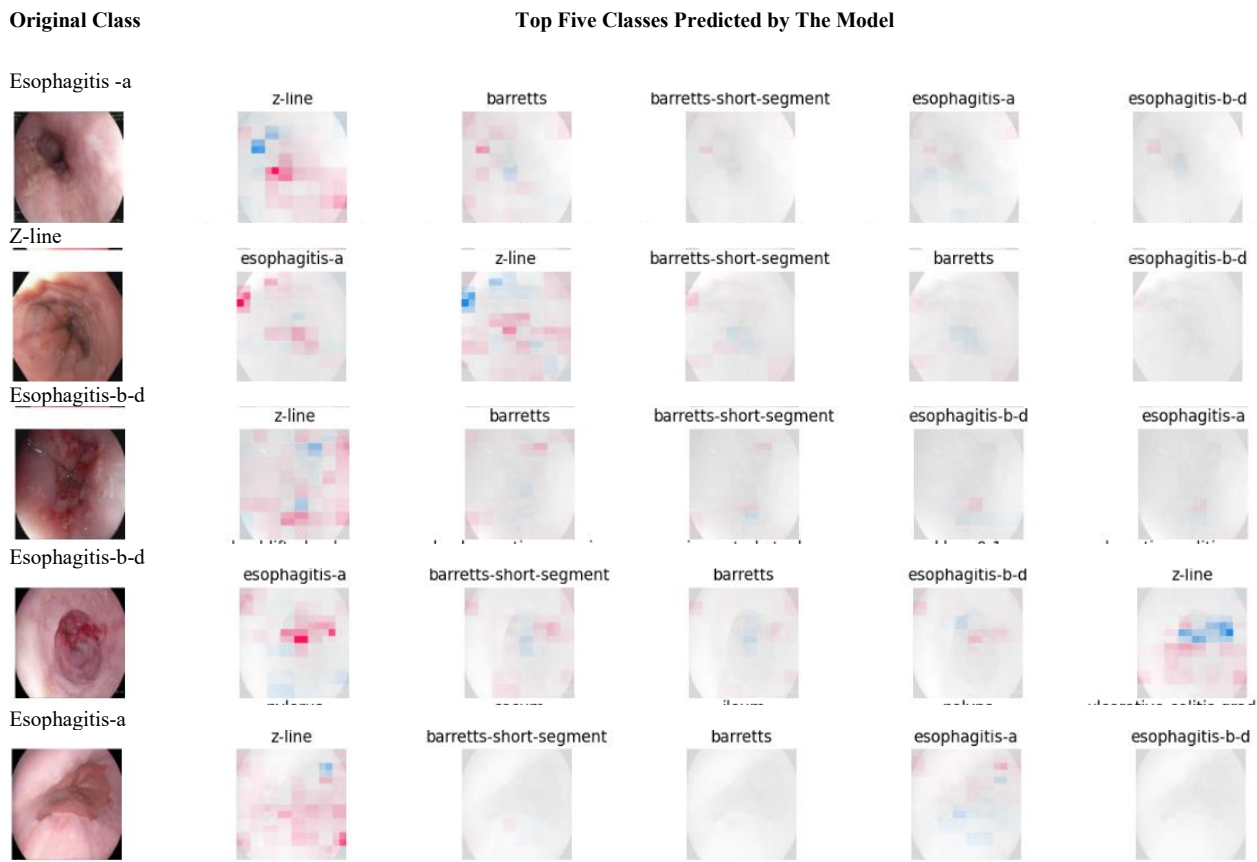

Figure 13. SHapley Additive exPlanations (SHAP) for misclassified images for the proposed model

As shown in Figure 14, a number of test samples of photos were chosen at random to provide the XAI predictions. For example, the sample image in the first row is part of the dyed resection margin class. The heatmap more accurately shows the impacted area, while the heatmap visualizes the area that is critical for class identification. Surprisingly, the Grad-CAM & SHAP incorporate the same area of the original image, exposing particular elements important for class prediction. Similarly, a sample image of an esophagitis-b-d is displayed in the fourth row where GradCAM and SHAP indicate the same area crucial for prediction. The visualizations show that the Grad-CAM precisely focusses on the specific class, and the heatmap detects the disease area more precisely. A sample image of an impacted stool class is displayed in the last row where GradCAM and SHAP indicate the same area important for classifying the exact class both indicated by red pixels. The figure demonstrates that heatmap, Grad CAM and SHAP visualization on a specific image of a class are indicating the same area important for visualization. Each sample is marked by red circle in the original image which is found by Grad CAM & SHAP, and the red circle presents the crucial portion responsible for classification. XAI with Grad-CAM and SHAP offers critical insights into "black box" AI models. Grad-CAM produces heatmaps that demonstrate which areas of an image a model focused on for a diagnosis (for example, identifying a suspicious lesion on an X-ray). SHAP assigns an importance value to each input feature, indicating, for example, whether patient symptoms or lab data had the most impact on disease prediction. These visualisations help physicians by increasing trust in AI judgements, validating findings (for example, confirming that the AI focused on the clinically relevant area), and perhaps identifying novel diagnostic markers, thereby supporting a collaborative human-AI approach to patient care.

## F. DISCUSSION
### 1) COMPARISON WITH LITERATURE

Table 15 compares the proposed framework with prior studies on GI disease classification using Hyperkvasir and other datasets, summarizing models and methods from various references. A variety of cutting-edge CNN models and their combinations are used in the deep learning architecture that have been proposed thus far for the categorization of gastrointestinal diseases. The majority of these models use certain frameworks to enhance the dataset's overall training and loss. However, there is still potential for both general and class-specific accuracy to improve. Hmoud *et al.* [39] used the Kvasir dataset with 5000 images and 5 classes. The AlexNet model was employed, achieving 97% accuracy. No information on F1, precision, recall, or explainability methods. Ramzan *et al.* [40] employed InceptionNetV3 and GITNet with QSVM utilizing the Kvasir dataset with 4000 images and 5 classes, achieving 99.32% accuracy. Grad-CAM was used for explainability. Khan *et al.*[41] employed DarkNet 53 and Xception (ESKNN), achieving 98.25% accuracy. Abraham *et al.*[42]used the Kvasir dataset with 5000 images and 5 classes.He employed Custom CNN with EfficientNetB0 with 5.3M parameters, achieving 98.01% accuracy, 98% with 5.3M parameters, achieving 98.01% accuracy, 98% precision, and 98% recall. Grad-CAM was used for explainability.

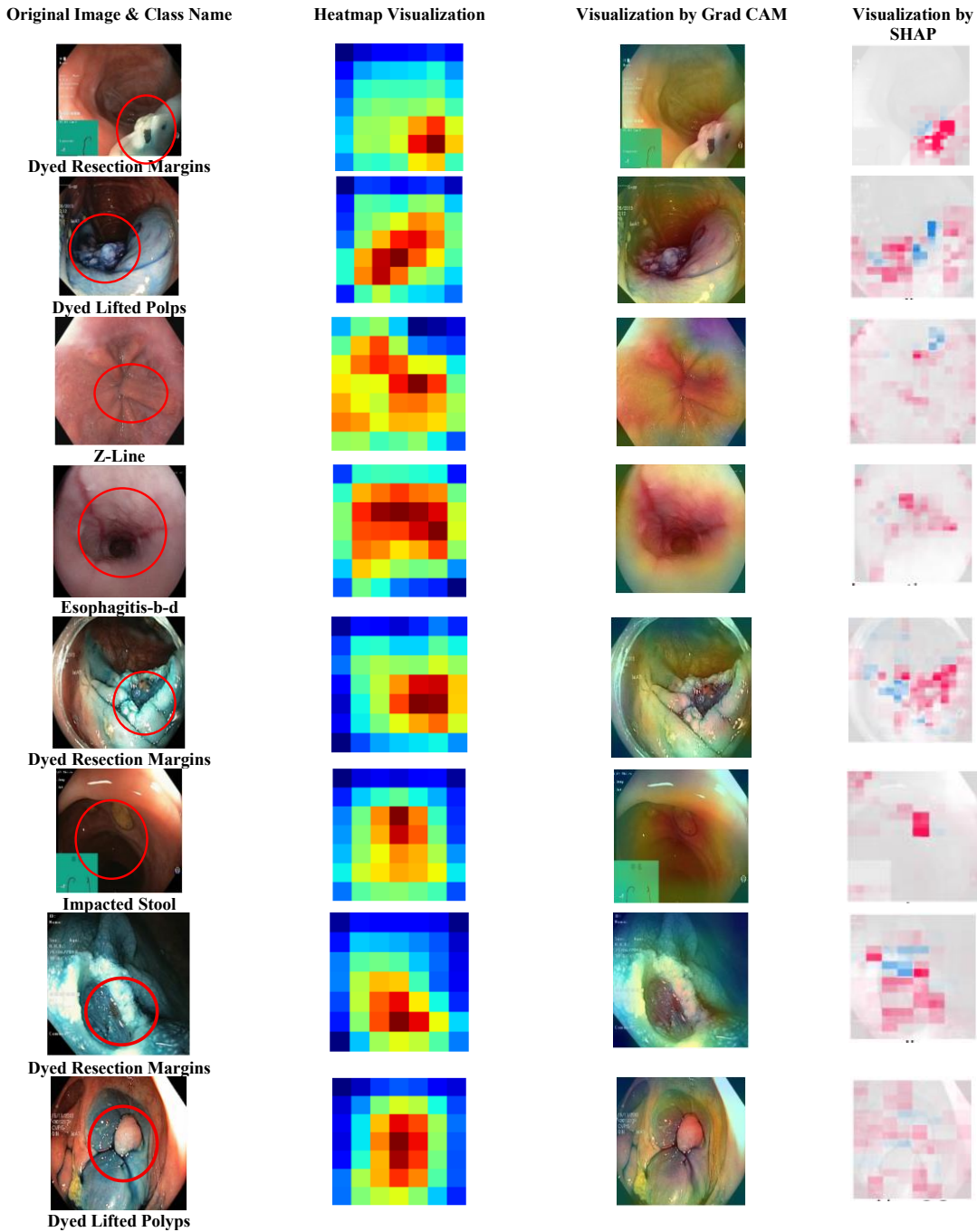

Figure 14. Heatmap, GradCAM & SHAP visualization on the same image of a specific class

Table 15. Results of previous studies compared with the proposed GastroViT model

| Ref. | Dataset | No. of sample images | No of class | Model | Parameters | Accuracy | F1 Score | Precision | Recall | XAI |
|---|---|---|---|---|---|---|---|---|---|---|
| Hmoud et al.[39] | Kvasir | 5000 | 5 | AlexNet | 62.3M | 97% | - | - | - | No |
| Ramzan et al. [40] | Kvasir | 4000 | 5 | InceptionNetV3, GITNet with QSVM | | 99.32% | | | | Grad CAM |
| Khan et al.[41] | KvasirV2 | 8000 | 8 | DarkNet 53,Xception(ESKNN) | - | 98.25% | - | - | - | No |
| Abraham et al.[42] | Kvasir | 5000 | 5 | Custom CNN with EfficintNetB0 | 5.3M | 98.01% | | 98% | 98% | Grad CAM |
| Öztürk et al. [26] | Kvasir | 6000 | 8 | Residual LSTM Layered ResNet50 | 23.9M | 98.05% | 98.05% | 98.05% | - | No |
| Ahamed et al.[26] | GastroVision KvasirV2 | 8000 8000 | 27 8 | PD-CNN | 0.815M | 87.75% 98.01% | 87.12% | 88.12% | 87.75% | Grad CAM, SHAP |
| Ramamurthy et al.[5] | Hyperkvasir | 10662 | 23 | Feature Fusion (EfficintNetB0 +Effmix) | 7.04M | 97.99% | 97% | 97% | 98% | No |
| Espantaleón-Pérez et al. [43] | Hyperkvasir | 10662 | 23 | MobileViT+ MobileViT_Large (Late Fusion) | 20M | - | 63.6% | - | - | No |
| Sarsengeldin et al, [44] | Kvasir Hyperkvasir | 8000 10662 | 5 23 | VGG16+CapsNets | - | 91% 85% | 92% 55% | 91% 54% | - | No |
| Yue et al, [45] | Custom Dataset Hyperkvasir | 22935 10662 | 2 23 | MobileNetV2 | - | 94.28% 90.44% | 90.40% 65.42% | - | - | No |
| Wang et al, [46] | KvasirV2 Hyperkvasir | 6000 6703 | 6 18 | L-DenseNetCaps | - | 94.83% 85.99% | 94% – | - | - | Heatmap |
| **Proposed GastroViT** | **Hyperkvasir** | **10662** | **23** | **Average Ensemble of MobileViT_XS & MobileViT_V2_200** | **20M** | **91.98%** | **64%** | **69%** | **63%** | **GradCAM SHAP** |

Öztürk et al. [26] utilized the Kvasir dataset with 6000 images and 8 classes and employed a Residual LSTM Layered ResNet50, achieving 98.05% accuracy, F1, precision, and recall. No explainability methods used. It is challenging to understand these models' conclusions or guarantee their clinical reliability because they rely on comparatively limited datasets, which may restrict generalization, and the majority of them lack explainability approaches. Complex systems like Residual LSTM, Xception, and DarkNet53 also make deployment more challenging and raise computing costs. Md. F. Ahamed et al.[26] used GastroVision and KvasirV2 datasets with 8000 images each and employed PD-CNN with 0.815M parameters, achieving 87.75% accuracy (GastroVision) and 98.01% accuracy (KvasirV2), with corresponding F1, precision, and recall values. Grad-CAM and SHAP were used for explainability. R. Espantaleón-Pérez et al.[43] used the Hyperkvasir dataset with 10662 images and 23 classes and employed MobileViT and MobileViT_Large (Late Fusion), achieving a precision of 63.6%. No information on F1, recall, or explainability methods. M. Sarsengeldin et al.[44] used Kvasir and Hyperkvasir datasets with 8000 and 10662 images respectively, having 5 and 23 classes and employed VGG16 + CapsNets, achieving 91% (Kvasir) and 85% (Hyperkvasir) accuracy, with varying precision, recall, and F1 scores. No explainability methods used. Wang et al.[46] used KvasirV2 and Hyperkvasir datasets with 6000 and 6703 images respectively, having 6 and 18 classes and employed L-DenseNetCaps, achieving 94.83% (KvasirV2) and 85.99% (Hyperkvasir) accuracy. Heatmap was used for explainability. These methods often show lower performance on larger or more complex datasets, with MobileViT achieving only 63.6% precision, and CapsNet-based models struggling on HyperKvasir. Additionally, most lack comprehensive explainability, limiting interpretability and clinical trust, while complex architectures increase computational requirements. Finally, the proposed model GastroViT is presented with the help of interpretability techniques as GradCAM and SHAP, the proposed GastroViT model achieved a high accuracy of 91.98% with F1 score 64%, Precision 69% and recall 63%on the Hyperkvasir dataset utilizing an ensemble of MViT_XS and MViTv2_200. The fact that the current study used a sizable dataset with 23 distinct classes is the main reason for this decreased classification accuracy. Furthermore, the HyperKvasir dataset has a variety of images depicting various gastrointestinal disorders. The scaling of images during the preprocessing stage, which causes a considerable loss of resolution, is another problem that affects the classification accuracy. Additionally, there are extremely few sample photos in some classes of the dataset, like ileum (only 9 sample images), haemorrhoids (6 sample images), and ulcerative colitis-grade 1-2(11 sample images). As a result, the model still has room for further development. The authors' future work will concentrate on balancing the HyperKvasir dataset in order to increase the model's efficiency. To improve categorization findings, the researchers intend to collect and produce a dataset that is more evenly distributed.

2)STRENGTHS, LIMITATIONS AND FUTURE WORK

With only 20M parameters, the MobileViT ensemble model obtained a comparable accuracy of 91.98% across 23 classes. During testing times of 0.007s and 0.003s, the suggested model—an ensemble of MViT_XS and MViTv2_200—achieved remarkable accuracy for all 23 classes. This study shows the efficiency of ViT models than the traditional CNN based models. By merging predictions from several models and applying a straightforward averaging technique, an average ensemble increases the accuracy and resilience of the model. Average ensembles provide a useful, effective means of achieving more dependable and precise results across a range of tasks by utilizing the strengths of diverse models, even weaker ones. Unlike other approaches, the suggested framework uses a very small number of parameters (20 million), which dramatically lowers the amount of computer resources needed. Due to its efficiency, the model can be deployed in situations with limited computational resources with more accessibility and practicality. Furthermore, the framework integrates explainable artificial intelligence (XAI) techniques like Grad-CAM and SHAP to emphasize interpretability. Hyperparameter tuning optimizes the model's design. As demonstrated, the model outperforms the SOTA-TL models in terms of classification performance and computing needs. With adequate accuracy, this method shortens testing times, drops layers, and uses fewer parameters. The suggested model is now easier to understand because it concentrates on pertinent image regions in order to extract valuable features, as demonstrated by the employment of Grad-CAM, Heatmap, and SHAP. Most previous studies have demonstrated their proposed models using these datasets, classifying 5 to 8 types of GI diseases. However, Hyperkvasir contains 23 different GI tract diseases. Additionally, baseline results have been established on this dataset for GI disease detection and classification of the upper, lower, and combined GI tract, offering valuable research resources for advancing GI endoscopy studies. Although the suggested method yields outcomes comparable to those of a lightweight model, it has several drawbacks. The model's classification accuracy is inferior to that of other existing studies. The primary cause of this lower classification accuracy is that the current study worked on a large dataset with 23 different classes. Additionally, the Hyperkvasir dataset has diverse types of images of different GI diseases. Moreover, some classes of the dataset contain very few sample images, such as hemorrhoids, which has only 6 sample images, and ileum, which has 9 sample images.

Future endeavors of the authors will focus on improving the model's efficiency by balancing the Hyperkvasir dataset. Future research will enhance gastrointestinal disease classification by employing advanced data augmentation, synthetic data generation, and interpretable deep learning models. Clinical validation, integrating multimodal data, and using unsupervised learning methods for pretraining unlabeled data will be key focuses. Furthermore, a variety of picture formats depicting various GI disorders may be found in the Hyperkvasir collection. The model still has room for further development. Subsequent investigations will improve the classification of gastrointestinal diseases by utilizing deep learning models that can be interpreted, enhanced data augmentation, and synthetic data synthesis. Important areas of study will include pretraining on unlabeled data utilizing unsupervised learning techniques, integrating multimodal data, and clinical validation.

## V. Conclusion

This study presents GastroViT, a vision transformer-based ensemble framework designed to improve the classification of GI diseases by addressing the limitations of conventional CNNs. Leveraging the self-attention mechanisms of MViT_XS and MViTv2_200, the proposed model effectively captures global characteristics in GI images. By employing an average ensemble of predictions from these two lightweight transformers, GastroViT achieves high classification performance across up to 23 classes, with macro-average precision, recall, F1 score, and testing accuracy reaching 69±0.36%, 63±0.39%, 64±0.38%, and 91.98%, respectively. When classes with fewer images were removed, performance further improved, achieving precision, recall, F1, and accuracy of 87±0.13%, 86±0.15%, 87±0.15%, and 92.70% for 16 classes. The study also demonstrates that GastroViT is computationally efficient, requiring only 20 million parameters, with low average testing times (0.01s per image for the ensemble), making it suitable for real-time clinical applications. Comparative analyses with transfer learning and state-of-the-art models highlight the superior performance of the proposed framework. Additionally, the integration of explainable AI methods such as Grad-CAM and SHAP provides interpretability, allowing clinicians to better understand model predictions. Overall, the work shows that GastroViT is an accurate, robust, and interpretable framework for GI disease classification, offering a promising tool for early diagnosis. Future directions include handling class imbalance and enhancing model robustness through further architectural improvements.

**Appendix:**

Table 10. Classwise Performance among modified transfer learning models

| Class Name | VGG16 | | | DenseNet201 | | | EfficientNetB7 | | | ResNet50 | | |
|---|---|---|---|---|---|---|---|---|---|---|---|---|
| | Precision | Recall | F1 Score | Precision | Recall | F1 Score | Precision | Recall | F1 Score | Precision | Recall | F1 Score |
| barretts | 0.00 | 0.00 | 0.00 | 0.00 | 0.00 | 0.00 | 1.00 | 0.11 | 0.20 | 0.00 | 0.00 | 0.00 |
| barretts-short | 0.00 | 0.00 | 0.00 | 0.00 | 0.00 | 0.00 | 0.00 | 0.00 | 0.00 | 0.00 | 0.00 | 0.00 |
| bbps-0-1 | 0.99 | 0.97 | 0.98 | 0.98 | 0.95 | 0.96 | 0.98 | 0.96 | 0.97 | 0.98 | 0.96 | 0.97 |
| bbps-2-3 | 0.98 | 0.98 | 0.98 | 0.97 | 0.98 | 0.97 | 0.97 | 0.98 | 0.98 | 0.98 | 0.99 | 0.98 |
| cecum | 0.95 | 0.98 | 0.97 | 0.98 | 0.99 | 0.98 | 0.96 | 0.96 | 0.96 | 0.98 | 0.99 | 0.98 |
| dyed-lifted polyps | 0.83 | 0.85 | 0.84 | 0.91 | 0.88 | 0.89 | 0.90 | 0.85 | 0.87 | 0.98 | 0.96 | 0.97 |
| dyed-resection margins | 0.85 | 0.82 | 0.84 | 0.88 | 0.91 | 0.90 | 0.86 | 0.90 | 0.88 | 0.86 | 0.86 | 0.86 |
| esophagitis-a | 0.46 | 0.48 | 0.47 | 0.51 | 0.44 | 0.48 | 0.48 | 0.44 | 0.46 | 0.51 | 0.46 | 0.48 |
| esophagitis-b-d | 0.72 | 0.56 | 0.63 | 0.74 | 0.65 | 0.69 | 0.65 | 0.58 | 0.61 | 0.71 | 0.56 | 0.62 |
| hemorrhoids | 0.00 | 0.00 | 0.00 | 0.00 | 0.00 | 0.00 | 0.00 | 0.00 | 0.00 | 0.00 | 0.00 | 0.00 |
| ileum | 0.00 | 0.00 | 0.00 | 0.00 | 0.00 | 0.00 | 0.00 | 0.00 | 0.00 | 0.00 | 0.00 | 0.00 |
| impacted stool | 0.83 | 0.89 | 0.86 | 0.88 | 0.81 | 0.85 | 0.82 | 0.85 | 0.84 | 0.88 | 0.85 | 0.87 |
| polyps | 0.95 | 0.97 | 0.96 | 0.92 | 0.98 | 0.95 | 0.96 | 0.95 | 0.96 | 0.98 | 0.97 | 0.97 |
| pylorus | 0.98 | 0.99 | 0.98 | 0.99 | 0.99 | 0.99 | 0.97 | 0.99 | 0.98 | 0.98 | 1.00 | 0.99 |
| retroflex-rectum | 0.92 | 0.92 | 0.92 | 0.91 | 0.92 | 0.92 | 0.90 | 0.94 | 0.92 | 0.87 | 0.96 | 0.92 |
| retroflex stomach | 0.99 | 0.98 | 0.98 | 1.00 | 0.99 | 0.99 | 0.99 | 0.99 | 0.99 | 0.99 | 0.97 | 0.98 |
| ulcerative-colitis-grade-0-1 | 0.00 | 0.00 | 0.00 | 0.50 | 0.14 | 0.22 | 0.00 | 0.00 | 0.00 | 0.25 | 0.14 | 0.18 |
| ulcerative-colitis-grade-1 | 0.58 | 0.44 | 0.50 | 0.61 | 0.27 | 0.37 | 0.37 | 0.34 | 0.35 | 0.57 | 0.41 | 0.48 |
| ulcerative-colitis-grade-1-2 | 0.00 | 0.00 | 0.00 | 0.00 | 0.00 | 0.00 | 0.00 | 0.00 | 0.00 | 0.00 | 0.00 | 0.00 |
| ulcerative-colitis-grade-2 | 0.60 | 0.72 | 0.65 | 0.61 | 0.84 | 0.71 | 0.60 | 0.73 | 0.66 | 0.62 | 0.75 | 0.68 |
| ulcerative-colitis-grade-2-3 | 0.00 | 0.00 | 0.00 | 0.00 | 0.00 | 0.00 | 0.00 | 0.00 | 0.00 | 0.00 | 0.00 | 0.00 |
| ulcerative-colitis-grade-3 | 0.62 | 0.67 | 0.64 | 0.64 | 0.56 | 0.61 | 0.73 | 0.59 | 0.65 | 0.56 | 0.67 | 0.61 |
| z-line | 0.74 | 0.85 | 0.79 | 0.78 | 0.89 | 0.83 | 0.79 | 0.91 | 0.84 | 0.75 | 0.90 | 0.82 |
| Average±sd | 0.56±0.4075 | 0.57±0.4143 | 0.56±0.4094 | 0.60±0.3943 | 0.57±0.4169 | 0.58±0.4051 | 0.61±0.4026 | 0.57±0.4125 | 0.57±0.4041 | 0.58±0.4022 | 0.58±0.4153 | 0.58±0.4073 |
| AUC | 0.98 | | | 0.96 | | | 0.97 | | | 0.98 | | |
| Test Accuracy(%) | 86.55 | | | 88.16 | | | 87.02 | | | 87.50 | | |

Table 11. Classwise Performance among CNN-based pretrained models

| Class Name | VGG16 | | | DenseNet201 | | | EfficientNetB7 | | | ResNet50 | | |
|---|---|---|---|---|---|---|---|---|---|---|---|---|
| | Precision | Recall | F1 Score | Precision | Recall | F1 Score | Precision | Recall | F1 Score | Precision | Recall | F1 Score |
| barretts | 0.00 | 0.00 | 0.00 | 0.00 | 0.00 | 0.00 | 0.50 | 0.11 | 0.18 | 0.50 | 0.11 | 0.18 |
| barretts-short | 0.00 | 0.00 | 0.00 | 0.00 | 0.00 | 0.00 | 0.00 | 0.00 | 0.00 | 0.00 | 0.00 | 0.00 |
| bbps-0-1 | 0.94 | 0.95 | 0.94 | 1.00 | 0.89 | 0.94 | 0.97 | 0.94 | 0.95 | 0.97 | 0.96 | 0.97 |
| bbps-2-3 | 0.97 | 0.99 | 0.98 | 0.95 | 0.98 | 0.97 | 0.96 | 0.97 | 0.97 | 0.99 | 0.96 | 0.98 |
| cecum | 0.97 | 0.96 | 0.96 | 0.95 | 0.99 | 0.97 | 0.95 | 0.96 | 0.95 | 0.97 | 0.96 | 0.96 |
| dyed-lifted polyps | 0.79 | 0.75 | 0.77 | 0.95 | 0.63 | 0.76 | 0.87 | 0.84 | 0.86 | 0.89 | 0.83 | 0.86 |
| dyed-resection margins | 0.79 | 0.81 | 0.80 | 0.60 | 0.99 | 0.74 | 0.87 | 0.89 | 0.88 | 0.85 | 0.89 | 0.87 |
| esophagitis-a | 0.36 | 0.38 | 0.37 | 0.68 | 0.28 | 0.40 | 0.49 | 0.35 | 0.41 | 0.47 | 0.25 | 0.32 |
| esophagitis-b-d | 0.62 | 0.46 | 0.53 | 0.68 | 0.77 | 0.72 | 0.64 | 0.56 | 0.60 | 0.63 | 0.73 | 0.68 |
| hemorrhoids | 0.00 | 0.00 | 0.00 | 0.00 | 0.00 | 0.00 | 0.00 | 0.00 | 0.00 | 0.00 | 0.00 | 0.00 |
| ileum | 0.00 | 0.00 | 0.00 | 0.00 | 0.00 | 0.00 | 0.00 | 0.00 | 0.00 | 0.00 | 0.00 | 0.00 |
| impacted stool | 0.88 | 0.81 | 0.85 | 0.84 | 0.78 | 0.81 | 0.81 | 0.93 | 0.86 | 0.74 | 0.93 | 0.82 |
| polyps | 0.91 | 0.96 | 0.94 | 0.97 | 0.99 | 0.98 | 0.94 | 0.92 | 0.93 | 0.95 | 0.97 | 0.96 |
| pylorus | 0.96 | 0.95 | 0.96 | 0.99 | 0.99 | 0.99 | 0.96 | 0.97 | 0.97 | 0.99 | 0.97 | 0.98 |
| retroflex-rectum | 0.88 | 0.85 | 0.86 | 0.91 | 0.94 | 0.92 | 0.89 | 0.90 | 0.89 | 0.89 | 0.96 | 0.93 |
| retroflex stomach | 0.97 | 0.98 | 0.98 | 0.99 | 0.96 | 0.98 | 0.98 | 0.99 | 0.98 | 1.00 | 0.97 | 0.98 |
| ulcerative-colitis-grade-0-1 | 0.17 | 0.14 | 0.15 | 0.17 | 0.14 | 0.15 | 0.00 | 0.00 | 0.00 | 0.25 | 0.14 | 0.18 |
| ulcerative-colitis-grade-1 | 0.47 | 0.37 | 0.41 | 0.58 | 0.17 | 0.26 | 0.30 | 0.20 | 0.24 | 0.57 | 0.20 | 0.29 |
| ulcerative-colitis-grade-1-2 | 0.00 | 0.00 | 0.00 | 0.00 | 0.00 | 0.00 | 0.00 | 0.00 | 0.00 | 0.00 | 0.00 | 0.00 |
| ulcerative-colitis-grade-2 | 0.57 | 0.67 | 0.62 | 0.63 | 0.72 | 0.67 | 0.53 | 0.79 | 0.64 | 0.54 | 0.89 | 0.68 |
| ulcerative-colitis-grade-2-3 | 0.00 | 0.00 | 0.00 | 0.00 | 0.00 | 0.00 | 0.00 | 0.00 | 0.00 | 0.00 | 0.00 | 0.00 |
| ulcerative-colitis-grade-3 | 0.65 | 0.56 | 0.60 | 0.62 | 0.30 | 0.40 | 0.77 | 0.37 | 0.50 | 0.69 | 0.41 | 0.51 |
| z-line | 0.74 | 0.83 | 0.79 | 0.78 | 0.91 | 0.84 | 0.73 | 0.91 | 0.81 | 0.73 | 0.94 | 0.82 |
| Average±sd | 0.55±0.3911 | 0.54±03956 | 0.54±0.3930 | 0.58±0.3990 | 0.54±0.4222 | 0.54±0.4035 | 0.57±0.3907 | 0.55±0.4189 | 0.55±0.4026 | 0.59±0.3747 | 0.57±0.4231 | 0.56±0.3985 |
| AUC | 0.96 | | | 0.97 | | | 0.97 | | | 0.98 | | |
| Test Accuracy(%) | 83.19 | | | 84.89 | | | 85.51 | | | 86.60 | | |

Table 12. Classwise Performance among ViT-based pretrained models

| Class Name | ResNeXt50 | | | EfficientViT_B2 | | | CoAtNet0 | | |
|---|---|---|---|---|---|---|---|---|---|
| | Precision | Recall | F1 Score | Precision | Recall | F1 Score | Precision | Recall | F1 Score |
| barretts | 0.00 | 0.00 | 0.00 | 0.50 | 0.12 | 0.20 | 0.00 | 0.00 | 0.00 |
| barretts-short | 0.12 | 0.09 | 0.11 | 0.00 | 0.00 | 0.00 | 0.00 | 0.00 | 0.00 |
| bbps-0-1 | 0.98 | 0.96 | 0.97 | 0.95 | 0.98 | 0.97 | 1.00 | 0.77 | 0.87 |
| bbps-2-3 | 0.96 | 0.97 | 0.97 | 0.98 | 0.95 | 0.96 | 0.81 | 1.00 | 0.90 |
| cecum | 0.95 | 0.99 | 0.97 | 0.97 | 0.98 | 0.97 | 0.99 | 0.87 | 0.93 |
| dyed-lifted polyps | 0.92 | 0.93 | 0.92 | 0.92 | 0.91 | 0.92 | 0.86 | 0.91 | 0.88 |
| dyed-resection margins | 0.95 | 0.92 | 0.93 | 0.91 | 0.93 | 0.92 | 0.85 | 0.87 | 0.86 |
| esophagitis-a | 0.51 | 0.41 | 0.45 | 0.60 | 0.49 | 0.54 | 0.55 | 0.42 | 0.48 |
| esophagitis-b-d | 0.60 | 0.58 | 0.59 | 0.73 | 0.62 | 0.67 | 0.54 | 0.69 | 0.61 |
| hemorrhoids | 0.00 | 0.00 | 0.00 | 0.00 | 0.00 | 0.00 | 0.00 | 0.00 | 0.00 |
| ileum | 0.00 | 0.00 | 0.00 | 0.00 | 0.00 | 0.00 | 0.12 | 0.50 | 0.20 |
| impacted stool | 0.80 | 0.77 | 0.78 | 0.79 | 0.85 | 0.81 | 0.92 | 0.42 | 0.58 |
| polyps | 1.00 | 0.98 | 0.99 | 1.00 | 0.98 | 0.99 | 0.99 | 0.89 | 0.94 |
| pylorus | 0.99 | 0.99 | 0.99 | 0.98 | 0.98 | 0.98 | 0.74 | 0.96 | 0.83 |
| retroflex-rectum | 0.92 | 0.97 | 0.94 | 0.95 | 0.97 | 0.96 | 0.92 | 0.76 | 0.83 |
| retroflex stomach | 0.98 | 0.97 | 0.98 | 0.97 | 0.98 | 0.97 | 0.91 | 0.95 | 0.93 |
| ulcerative-colitis-grade-0-1 | 0.00 | 0.00 | 0.00 | 0.00 | 0.00 | 0.00 | 0.00 | 0.00 | 0.00 |
| ulcerative-colitis-grade-1 | 0.47 | 0.47 | 0.48 | 0.42 | 0.47 | 0.45 | 0.40 | 0.10 | 0.16 |
| ulcerative-colitis-grade-1-2 | 0.00 | 0.00 | 0.00 | 0.00 | 0.00 | 0.00 | 0.00 | 0.00 | 0.00 |
| ulcerative-colitis-grade-2 | 0.64 | 0.71 | 0.67 | 0.62 | 0.70 | 0.66 | 0.48 | 0.73 | 0.58 |
| ulcerative-colitis-grade-2-3 | 0.00 | 0.00 | 0.00 | 0.00 | 0.00 | 0.00 | 0.00 | 0.00 | 0.00 |
| ulcerative-colitis-grade-3 | 0.62 | 0.67 | 0.64 | 0.71 | 0.63 | 0.67 | 0.79 | 0.41 | 0.54 |
| z-line | 0.80 | 0.92 | 0.85 | 0.77 | 0.94 | 0.85 | 0.80 | 0.78 | 0.79 |
| Average±sd | 0.57±0.41 | 0.58±0.42 | 0.58±0.41 | 0.60±0.39 | 0.59±0.42 | 0.59±0.41 | 0.55±0.39 | 0.52±0.38 | 0.52±0.38 |
| AUC | 0.98 | | | 0.98 | | | 0.97 | | |
| Test Accuracy(%) | 88.79 | | | 89.02 | | | 81.85 | | |

The code for this research is given at https://github.com/Sumaiya-1810023/GastroViT_Code.

**Institutional Review Board Statement:** Not applicable

**Informed Consent Statement:** Not applicable

**Funding**
Not applicable

**Conflict of Interest**
The authors affirm that they do not possess any identifiable conflicting financial interests or personal affiliations that might have influenced the findings presented in this paper.

**Data Availability Statement**
The dataset of the current study can be made available through a reasonable request to the corresponding author.

**Acknowledgements**
Throughout the writing process, the authors improved the paper's language and readability using a variety of AI technologies. Following the use of these resources, the writers assessed and made any required edits to the text. The basic study, outcomes, and research findings are entirely the authors' responsibility.

The open-access publication of the article is supported by Qatar National Library (QNL).